\documentclass[largeformat]{interact}

\usepackage{epstopdf}% To incorporate .eps illustrations using PDFLaTeX, etc.
\usepackage[caption=false]{subfig}% Support for small, `sub' figures and tables

\usepackage{url}
\usepackage{hyperref}
\usepackage{caption}
\usepackage{tabularx}
\usepackage{graphicx}
\usepackage{subfig}
 \usepackage{xcolor}
 \usepackage{soul}

\usepackage[natbibapa,nodoi]{apacite}
\theoremstyle{plain}% Theorem-like structures provided by amsthm.sty

\theoremstyle{definition}

\theoremstyle{remark}

\begin{document}

\articletype{ARTICLE TEMPLATE}% Specify the article type or omit as appropriate

%\title{Taylor \& Francis \LaTeX\ template for authors (\textsf{Interact} layout + American Psychological Association reference style)}

\title{Interactive Explanation with Varying Level of Details in an Explainable Scientific Literature Recommender System}

%\title{Designing Interactive Explanation with Varying Level of Details in an Explainable Scientific Literature Recommender System: an HCD approach}

%\author{
%\name{Mouadh Guesmi\textsuperscript{a}\thanks{CONTACT Mouadh Guesmi. Email: mouadh.guesmi@stud.uni-due.de}, Mohamed Amine Chatti \textsuperscript{a}, Shoeb Joarder \textsuperscript{a}, Qurat Ul Ain \textsuperscript{a}, Rawaa Alatrash \textsuperscript{a}, Clara Siepmann \textsuperscript{a}, Tannaz Vahidi \textsuperscript{a}, Hoda Ghanbarzadeh \textsuperscript{a} and Jaleh Ghorbani-Bavani \textsuperscript{a}}
%\affil{\textsuperscript{a} University of Duisburg-Essen, Duisburg, Germany}}

\author{
\name{Mouadh Guesmi, Mohamed Amine Chatti\thanks{CONTACT Mohamed Amine Chatti. Email: mohamed.chatti@uni-due.de}, Shoeb Joarder, Qurat Ul Ain, Rawaa Alatrash, Clara Siepmann, and Tannaz Vahidi}
\affil{Social Computing Group, Faculty of Computer Science, University of Duisburg-Essen, Duisburg, Germany}}

\maketitle

\begin{abstract}
Explainable recommender systems (RS) have traditionally followed a one-size-fits-all approach, delivering the same explanation level of detail to each user, without considering their individual needs and goals. Further, explanations in RS have so far been presented mostly in a static and non-interactive manner. To fill these research gaps, we aim in this paper to adopt a user-centered, interactive explanation model that provides explanations with different levels of detail and empowers users to interact with, control, and personalize the explanations based on their needs and preferences. We followed a user-centered approach to design interactive explanations with three levels of detail (basic, intermediate, and advanced) and implemented them in the transparent Recommendation and Interest Modeling Application (RIMA). We conducted a qualitative user study (N=14) to investigate the impact of providing interactive explanations with varying level of details on the users’ perception of the explainable RS. Our study showed qualitative evidence that fostering interaction and giving users control in deciding which explanation they would like to see can meet the demands of users with different needs, preferences, and goals, and consequently can have positive effects on different crucial aspects in explainable recommendation, including transparency, trust, satisfaction, and user experience.
\end{abstract}

\begin{keywords}
Recommender system; Intelligent user interfaces; Explainable recommendation; Interactive explanation;
Intelligibility types; Visualization
\end{keywords}

\section{Introduction}
Research on Artificial Intelligence (AI) underscores the importance of user-friendly AI systems that foster trust, supported by factors like fairness, accountability, transparency, and ethics (FATE) \citep{choung2023trust,vianello2023improving,wahde2023daisy}. Crucial to trustworthy AI is the explainability of these systems, which has ushered a remarkable growth in techniques to
open the black-box of AI \citep{arrieta2020explainable,ehsan2021operationalizing}. Explainable AI (XAI) aims at making the decisions of AI systems comprehensible to affected humans \citep{adadi2018peeking,arrieta2020explainable}. The level of user comprehension not only influences their trust but also shapes their emotional confidence and interaction with the AI system \citep{shin2020user,shin2021effects,shin2022algorithm,shin2023algorithms}. Moreover, explanations are essential to hold AI systems accountable, and can serve as a means to ensure humans’ right to understand and object AI decisions \citep{ehsan2023human}.  

This interplay between user trust and explainability also forms a critical discourse in the Recommender Systems (RS) domain \citep{siepmann2023trust}. RS are widely used to make it simpler for users to find their desired products across many platforms among the vast variety of choices available. Nowadays, explainability represents a crucial feature of RS. Users may not understand the system's behavior without explanations, especially if the system behaves unexpectedly, causing a lack of confidence among users who may then lose trust, get frustrated, and eventually abandon the system. Explainable recommendation aims at making the RS more transparent and avoiding so-called “black-box” models. Generally, an explanation seeks to answer questions, also called intelligibility queries or types, such as \textit{What}, \textit{Why}, \textit{How}, \textit{What if}, \textit{What else}, and \textit{Why not} in order to achieve understanding \citep{lim2009assessing}.   

Previous Research on explainable AI (XAI) discussed several key characteristics of \textit{good} explanations. \cite{kulesza2013too} showed that explanations that are \textit{sound} and \textit{complete} can positively affect user understandability and trust. 
The concept of \textit{soundness} refers to the quality of providing only accurate and truthful information that increases in proportion to the extent to which these explanations accurately represent the underlying model. The concept of \textit{completeness} refers to the extent to which all of the underlying system is described by the explanation, concretely, the more the explanation describes the underlying system, the more complete the explanation is.
Balanced against the soundness and completeness principles is the need to remain comprehensible and to \textit{avoid overwhelming} users \citep{kulesza2015principles}. These principles imply that it is important to provide explanations with enough details to allow users to build accurate mental models of how the RS operates without overwhelming them. This is in line with other research showing that different users have different needs for explanation and explanations may cause negative effects (e.g., high cognitive load, confusion, lack of trust) if they are difficult to understand \citep{gedikli2014,kizilcec2016much,yang2020visual,zhao2019users}. Further, \cite{miller2019explanation} stressed that \textit{selective} and \textit{social} characteristics of explanation need to be taken into account in order to achieve meaningful explanation. The selective and social nature of explanation implies that an explainable RS has to be interactive. Understanding who interacts with the black-box of AI is just as important as opening it, if not more \citep{ehsan2022human}. To address this challenge, researchers have called for adopting human-centered approaches to XAI, which is inevitable given that explainability is a human-centric property and XAI must be studied as an interaction problem \citep{liao2021human,shin2023algorithms}. While many researchers recognize the necessity to provide interaction mechanisms in the context of explanations, there is still a lack of research on interactive explanation in RS. In particular, how to design and implement interactive explanation in RS, as well as its potential positive effect on e.g., transparency, trust, and satisfaction remain open questions \citep{jannach2019explanations,hernandez2021effects,schaffer2015hypothetical,guesmi2021demand}.

\cite{liao2021human} point out that what makes an explanation good is to provide appropriate information that can be
understood and utilized by the target user, which depends on the receiver’s current knowledge and their goal for receiving the explanation,
among other human factors. Despite the acknowledgment that the need for explanations may vary considerably between end-users, explainable RS have traditionally followed a one-size-fits-all model, whereby the same explanation level of detail is provided to each user, without taking into consideration individual user’s goals and personal characteristics \citep{chatti2022more,ain2022multi}. To address these research gaps, in this paper, we are particularly interested in how to systematically provide interactive explanation with varying level of details in an explainable RS. Our aim is to shift from a one-size-fits-all explainable recommendation approach toward a user-centered, interactive explanation model where users can steer the explanation process, change the explanation to answer intelligibility questions of interest (e.g., \textit{What}, \textit{What if}, \textit{Why}, and \textit{How} questions), and personalize the explanation based on their preferences (e.g., whether or not to see the explanation, see different levels of explanation detail, change the explanation viewpoint by focusing on the input, process, and/or output). Further, we conducted a qualitative user study (N=14) based on moderated think-aloud sessions and semi-structured interviews with students and researchers to investigate the potential effects of providing interactive explanation with varying level of details in an explainable RS. Due to the subjective nature of the required explanation levels of detail, a qualitative approach seems to be the most appropriate to investigate them. This approach allows us to investigate the users’ unique perspectives and expectations from an explainable RS in-depth and ask for what information should be contained at the different explanation levels of detail. 

To conduct this study, we designed explanations with three different levels of detail (basic, intermediate, advanced) and implemented them in the transparent Recommendation and Interest Modeling Application (RIMA) that provides interactive on-demand explanations of recommended scientific publications, in order to meet the needs and preferences of different users. The objective of the study was to answer the following research questions: Which and how much information should be provided at each explanation level of detail? \textbf{(RQ1)} Can providing interactive explanation with varying level of details positively affect the perception of the explainable RS in terms of (1) user control \& personalization, (2) transparency \& trust, (3) user satisfaction, and (4) user experience? \textbf{(RQ2)}. The results of our study show that there is qualitative evidence that following a user-centered, interactive approach has positive effects on different vital aspects in explainable recommendation, including transparency, trust, and satisfaction.

To summarize, this work makes the following three main contributions: First, we highlight the importance of providing interactive on-demand explanations with different levels of detail and empowering users to steer the explanation process. Second, we systematically design and implement explanations with three different levels of detail in an explainable RS. Third, we provide qualitative evidence on the potential positive impact of user-controlled, interactive explanation with varying level of details on the perception of explainable recommendation.  

%%%%%%%%%%%%%%%%%%%%%%%%%%%%%%%%%%%%%%%
\section{Related work}
\subsection{Explanation with different levels of detail}
The explanation level of detail is an important factor in the design process of explainable RS. In this work, the level of detail refers to the amount of information exposed in an explanation. In the field of XAI in general, many researchers, e.g., \cite{mohseni2018,miller2019explanation} argue that different user groups will have other goals in mind while using XAI systems. Therefore, the design choices in XAI should be driven by users’ explanation needs and goals \citep{chatti2022more,liao2021human,liao2023ai,ehsan2020human,shin2021effects}.  

Besides the goals of the users, another vital aspect that will influence their understanding of explanations are their cognitive capabilities. Results of previous research on XAI showed that for specific users or user groups, the detailed explanation does not automatically result in higher trust and user satisfaction because the provision of additional explanations increases cognitive effort \citep{kizilcec2016much, kulesza2015principles,zhao2019users, yang2020visual}. 
\cite{kulesza2015principles} outlined a set of principles for designing explanations to personalize interactive machine learning. These principles include "Be Sound", “Be Complete” and “Don’t Overwhelm” implying a trade-off between the amount of information in an explanation and the level of perceived transparency, trust, and satisfaction users develop when interacting with the AI system. \textit{Soundness} means telling nothing but the truth. It refers to the explanation fidelity, i.e., “the extent to which each component of an explanation’s content is truthful in describing the underlying system”. Evaluating soundness requires comparing the explanation with the learning system’s mathematical model, “the more these explanations reflect the underlying model, the more sound the explanation is”. \textit{Completeness} means telling the whole truth. It refers to “the extent to which all of the underlying system is described by the explanation”. A complete explanation informs users about all the information the learning system had at its disposal and how it used that information. The authors suggest that one method for evaluating completeness is via Lim and Dey's intelligibility types (e.g., input, model, why, what if, certainty) \citep{lim2009assessing}, with more complete explanations including more of these intelligibility types \citep{kulesza2015principles}. In an earlier study, \cite{kulesza2013too} concluded that there is a need to provide explanations with enough soundness and completeness in order to help users build an accurate mental model of how the system works without overwhelming them.

While the provision of explanations with varying level of details is gaining popularity in XAI research, its application in the field of explainable recommendation remains limited in the existing literature. To the best of our knowledge, only few studies have explored explanations that offer different levels of detail \citep{millecamp2019, guesmi2021open, guesmi2021input, guesmi2021demand, guesmi2022explaining, chatti2022more}. For instance, \cite{millecamp2019} developed a music recommendation system that not only enables users to choose whether or not to view explanations using a "Why?" button, but also allows them to select the level of detail through a "More/Hide" button. Similarly, \cite{chatti2022more} designed an explainable RS that provides on-demand personalized explanations for tweet recommendations, with three levels of detail to cater to the diverse requirements of different end-users. In another work, \cite{guesmi2022explaining} utilized the same system to explain black box user interest models with varying level of details, thereby facilitating transparent recommendations tailored to users' unique characteristics. While these works provided explanations at different levels of detail, the design of these explanations was not conducted in a systematic and theoretically-sound manner which represents a research gap that we aim to address in this work by building upon the principles of soundness and completeness introduced in \citep{kulesza2015principles} to define three explanation levels, and following a user-centered approach to design the explanation interfaces.  
%%%%%%%%%%%%%%%%%%%%%%%%%%%%%%%%%%%%%%%%%%
\subsection{Interactive explanation}  
User awareness and user control are key factors for successful user experience with personalization services, as evidenced by several recent studies \citep{sundar2020rise,shin2022algorithm,shin2021effects,liang2023promoting}. User awareness can be achieved by providing explanations about the how, why, and what an RS does, which can increase transparency and contribute to greater trust and positive user experience \citep{guesmi2023info,tintarev2015explaining,tsai2021effects}. Further, it is important to provide agency to users by enabling them to not only see explanations but also to control them \citep{guesmi2021demand}.

Explainability is an inherently human-centric property \citep{liao2021human}. In a broader view, an explanation aims to make the reasons behind a decision or recommendation easy to understand by people  \cite{arrieta2020explainable}. Thus, work on XAI in general and explainable RS in particular must take a human-centered approach. To this end, the HCI community has called for interdisciplinary collaboration and human-centered approaches to design, evaluate, and provide conceptual and methodological tools for XAI \citep{abdul2018trends, wang2019designing,liao2021human}. This has led to a sub-field of XAI referred to as human-centered XAI (HCXAI) which goes beyond opening the black-box and towards the human factors of XAI \citep{ehsan2020human}. This line of research is based on common assumptions that there is more to making AI explainable than algorithmic transparency and that who opens the black-box of AI matters just as much, if not more, as the ways of opening it \citep{ehsan2021operationalizing,ehsan2022human,ehsan2023human}. \cite{miller2019explanation} synthesized  perspectives on human explanation from philosophy, social science, and cognitive science and identified a list of human-friendly characteristics of explanation, including that human explanations are contrastive (i.e., “sought in response to particular counterfactual cases”), selective (i.e., selected in a ‘‘biased manner” from a “sometimes infinite number of causes”), and social (i.e., conversational process, where an "explainer transfer knowledge to an explainee"). 
The social nature of explanation maps to an essential requirement for interactivity in XAI applications \citep{liao2020questioning}. User interactions do not end at receiving an XAI output, but continue until an actionable understanding is achieved \citep{liao2021human}.
With the goal of bridging the gap between XAI and HCI, research on designing and studying user interactions with XAI has emerged over the past few years \citep{cheng2019explaining, sokol2020one, krause2016interacting, kulesza2015principles}. 
However, little is known about how interactive explanation should be designed and implemented in RS, so that explanation goals such as scrutability, transparency, trust, and user satisfaction are met \citep{jannach2019explanations, hernandez2021effects}. Although interactive and more recently conversational RS have been well studied \citep{he2016interactive, jin2018effects, jugovac2017interacting, harambam2019designing, jannach2021survey}, there has been little work on how to incorporate interactivity features in explainable RS. Both in the literature and in real-world systems, there are only a few examples of RS that provide interactive explanations, mainly to allow users to scrutinize the provided recommendations and correct the system’s assumptions \citep{jannach2019explanations, guesmi2021open, guesmi2022interactive, guesmi2021input, guesmi2022explaining, balog2019transparent}, or have a conversation, i.e., an exchange of questions and answers between the user and the system, using GUI-navigation or natural language conversation \citep{hernandez2021effects}. As a possible mechanism to achieve interactive explanation in RS, we focus in this work on developing explanations with different levels of detail and empowering users to steer the explanation process the way they see fit.

%%%%%%%%%%%%%%%%%%%%%%%%%%%%%%%%%%%%%%%%%%%%%%%%%%
\section{RIMA application}
In this work, we focus on recommending scientific publications and leveraging explanatory visualizations to provide interactive explanations with varying level of details. To conduct our study, we designed explanations with three different levels of detail (basic, intermediate, advanced) and implemented them in the transparent Recommendation and Interest Modeling Application (RIMA), which a content-based RS that produces on-demand content-based explanations \citep{guesmi2021open,guesmi2021demand,guesmi2021input,chatti2022more,guesmi2022explaining}. RIMA was developed with the purpose of autonomously extracting users' interests from their previous scientific publications, which are subsequently utilized to provide publication recommendations relevant to the users' areas of focus 
\citep{guesmi2022if,guesmi2022interactive,guesmi2023info,guesmi2023validation}. 

\subsection{Interest model generation}
The user interest models in RIMA are automatically inferred from users’ publications. We utilized the \textit{SIFRank} \citep{sifrank} technique, which is an embedding-based approach for extracting keyphrases, to extract keyphrases from an author's published works. Our choice of SIFRank was motivated by its outstanding performance, as it achieved state-of-the-art results, surpassing other unsupervised keyphrase extraction methods that rely on pre-trained language models. Compared to SIFRank's performance when using popular transformer models like BERT, RoBERTa, and XLNet, the SIFRank's developers found that using ELMo for word embeddings produced better results \citep{sifrank}. However, it's important to note that ELMo, being an LSTM-based approach, can be computationally intensive \citep{yu2020bert}. Therefore, we replaced the ELMo word embedding method for SIFRank with the pretrained transformer model \textit{SqueezeBERT} \citep{squeezebert}. This decision was based on its enhanced inter-layer information flow and the fact that its transformer design is lightweight, making it 4.3 times faster than the BERT model \citep{squeezebert}.

Leveraging a knowledge base in the process of inferring interest models has the potential to resolve various semantic-related problems, such as merging synonym interests, reducing acronym interests, and eliminating noise caused by irrelevant keyphrases \citep{chatti2021simt}. Consequently, this leads to the creation of more comprehensive and precise interest models. In this work, we employed \textit{DBpedia Spotlight} \citep{dbpedia-spotlight} as an entity linking service to establish connections between keyphrases and concepts within the DBpedia knowledge base \citep{dbpedia}. This allowed us to create user interest models that are enriched with semantic information. Based on these inferred interest models, the recommendation engine provides scientific publication recommendations. The top five interests, determined by their weights, are initially used as input for the recommendation process. 

\subsection{Recommendation generation}
For obtaining the candidate publications, we use the semantic scholar API to fetch publications that contain, or are related to, one or more user interests, which serve as input for the recommendation system. We then apply \textit{SingleRank} \citep{singlerank}, an unsupervised keyphrase extraction algorithm, on the fetched publications to extract keywords from the title and the abstract text. After that, we represent the user's interest model and the keyphrases extracted from the collected publications as embedding vectors. For this purpose, we used the sentence transformer model \textit{msmarco-distilbert-base-tas-b}\footnote{https://huggingface.co/sentence-transformers/msmarco-distilbert-base-tas-b} to extract semantic representations of users models and candidate publications, enabling us to measure and capture the semantic similarity between them.
We used the cosine similarity measure to obtain a semantic similarity score. The top ten similar publications will be then recommended to the user. Initially, publications exceeding a 40\% semantic similarity threshold are presented to the user.

%%%%%%%%%%%%%%%%%%%%%%%%%%%%%%%%%%%%%%%%%%%%%%%%%%%%%%%%%%%%%%%%%%%%%%%%%%%%%%%%
\section{Explanation design}
In this section, we present in detail the steps that we followed to systematically design the explanations with three different levels of detail in RIMA. We first present the building blocks of the explanations based on the \textit{What}, \textit{What if}, \textit{Why}, and \textit{How} intelligibility types. We then explain how we combined these intelligibility types to provide basic, intermediate, and advanced explanations. 

\subsection{Intelligibility types} \label{Int_Types}
Being able to “represent to their users what they know, how they know it, and what they are doing about it” is what it means for an application to be deemed \textit{intelligible} \citep{bellotti2001intelligibility}. In our work, we use Lim and Dey’s popular categorization of intelligibility types \citep{lim2009assessing} to systematically design explanation with varying level of details. Specifically, we are interested in \textit{What}, \textit{What if}, \textit{Why}, and \textit{How} intelligibility types, presented in Table \ref{table:1} with their addressed questions. 

\begin{table}[h]
\centering
\begin{tabular}{ |c|c| } 
 \hline
    Intelligibility types & Addressed questions \\
 \hline
 \textit{What} & What information does the system have?  \\ 
 \hline
 \textit{What} if & What would happen if the inputs were changed? \\
 \hline
 \textit{Why} & Why is this publication recommended?  \\
 \hline
 \textit{How} & How did the system generate this recommendation?  \\
 \hline
\end{tabular}
\caption{Intelligibility types.}
\label{table:1}
\end{table}

\begin{figure}[h]
     \centering
     \subfloat[What.]{
        \includegraphics[width=0.48\textwidth]{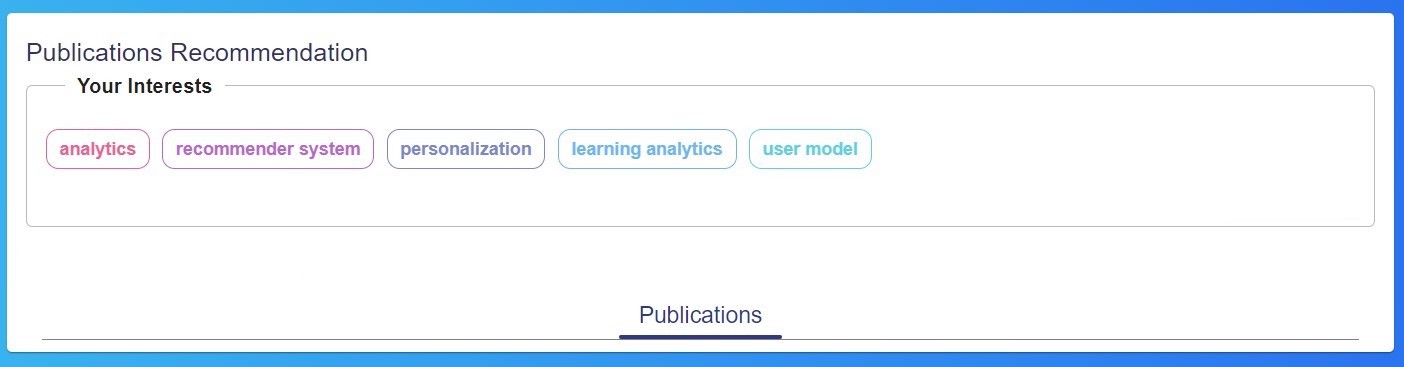}
        \label{fig:what_int}
     } 
     
     \hfill   
     
     \subfloat[What if.]{
        \includegraphics[width=0.48\textwidth]{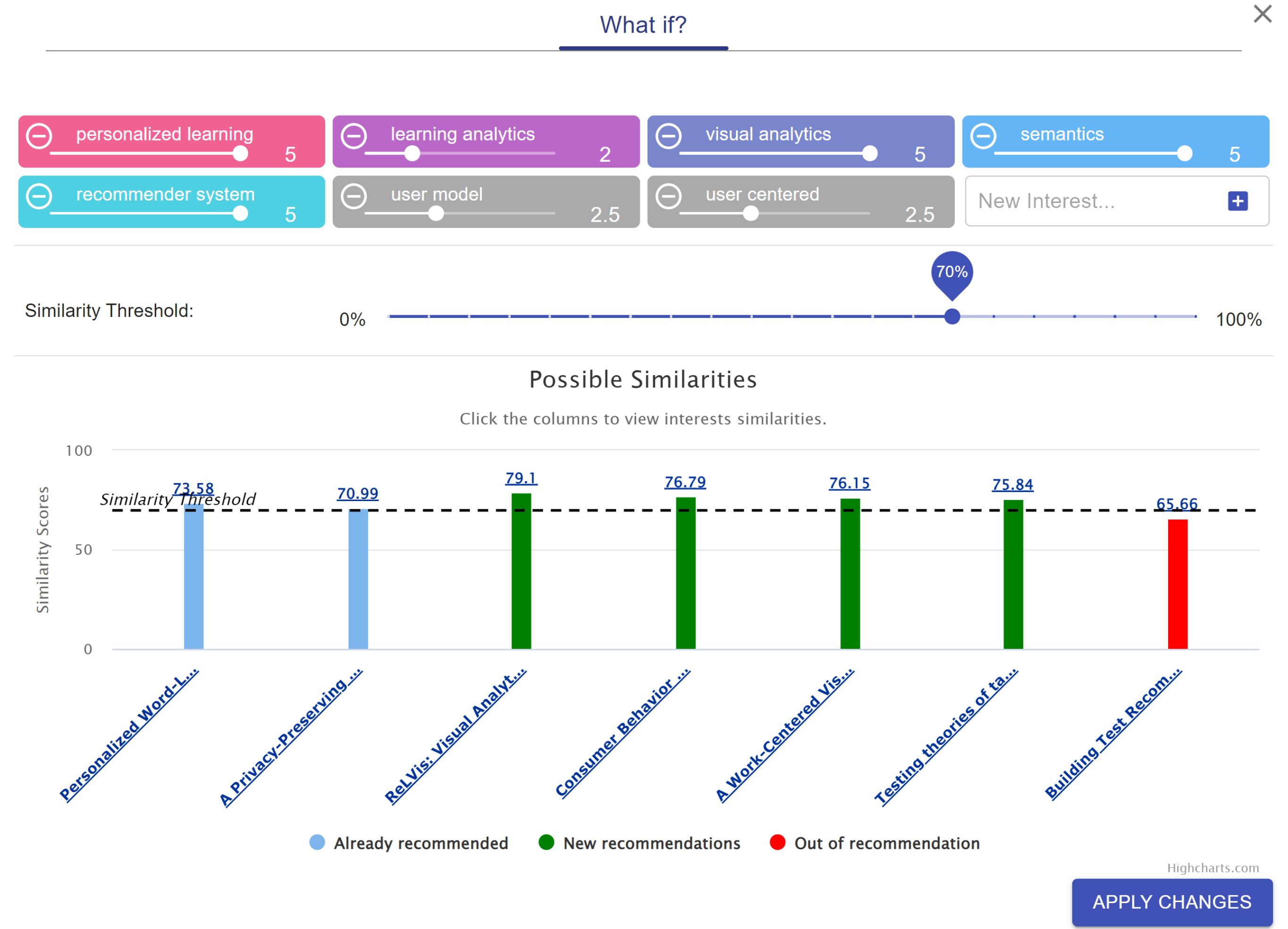}
        \label{fig:whatif_int}
     }
        \caption{What and What-if intelligibility types.}
        \label{fig:IT1}
\end{figure}

\begin{figure}[h]
     \centering
     \subfloat[Why (abstract).]{
        \includegraphics[width=0.48\textwidth]{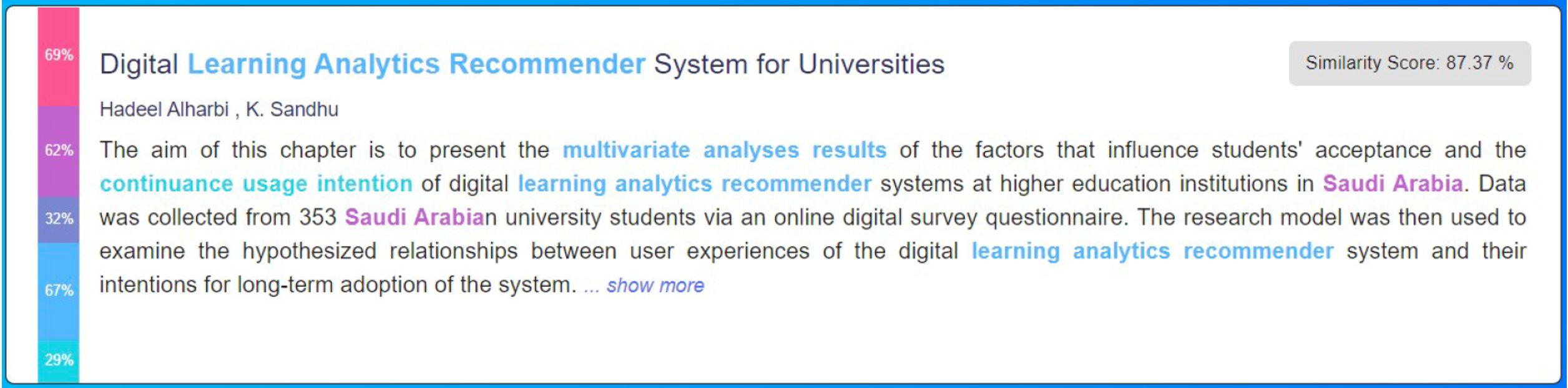}
        \label{fig:why_abstract}
   }
    \hfill
    \subfloat[Why (detailed).]{
         \includegraphics[width=0.42\textwidth]{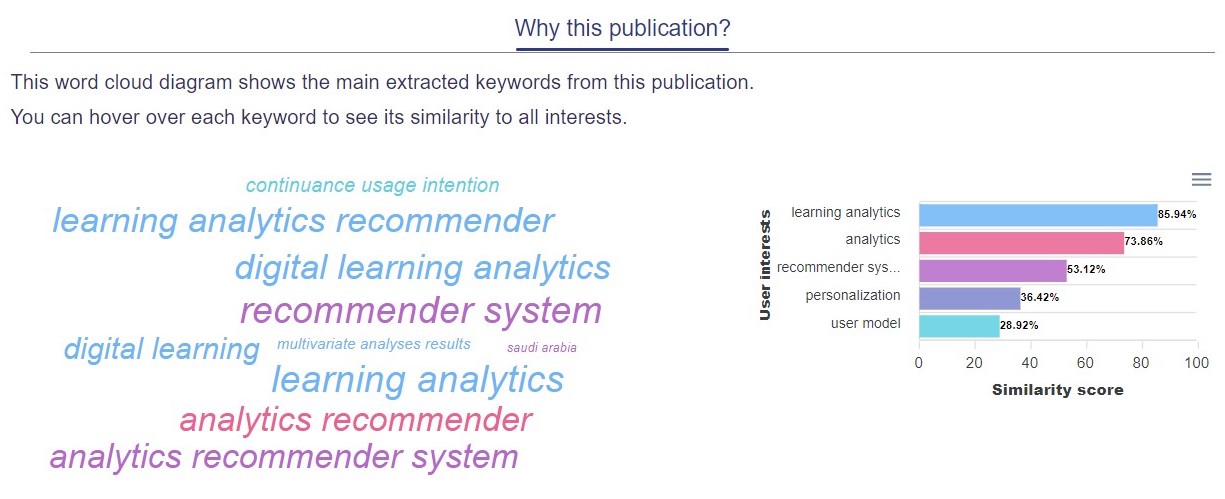}
         \label{fig:why_detail}
     }
        \caption{Why (abstract and detailed) intelligibility types.}
        \label{fig:IT2}
\end{figure}

\begin{figure}[h]
     \centering
     \subfloat[How – overview.]{
         \includegraphics[width=0.6\textwidth]{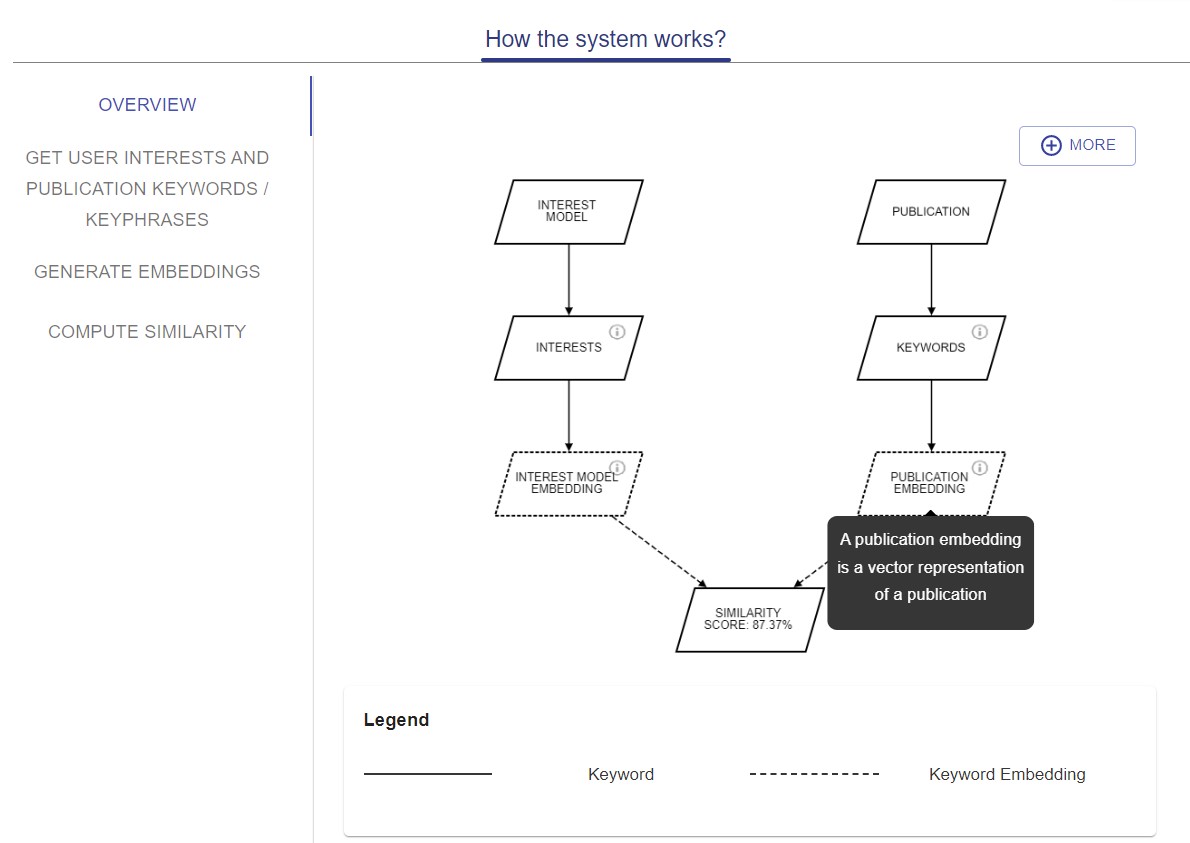}
         \label{fig:how_overview}
     }
     
     \hfill
     
     \subfloat[How – detailed.]{
         \includegraphics[width=0.7\textwidth]{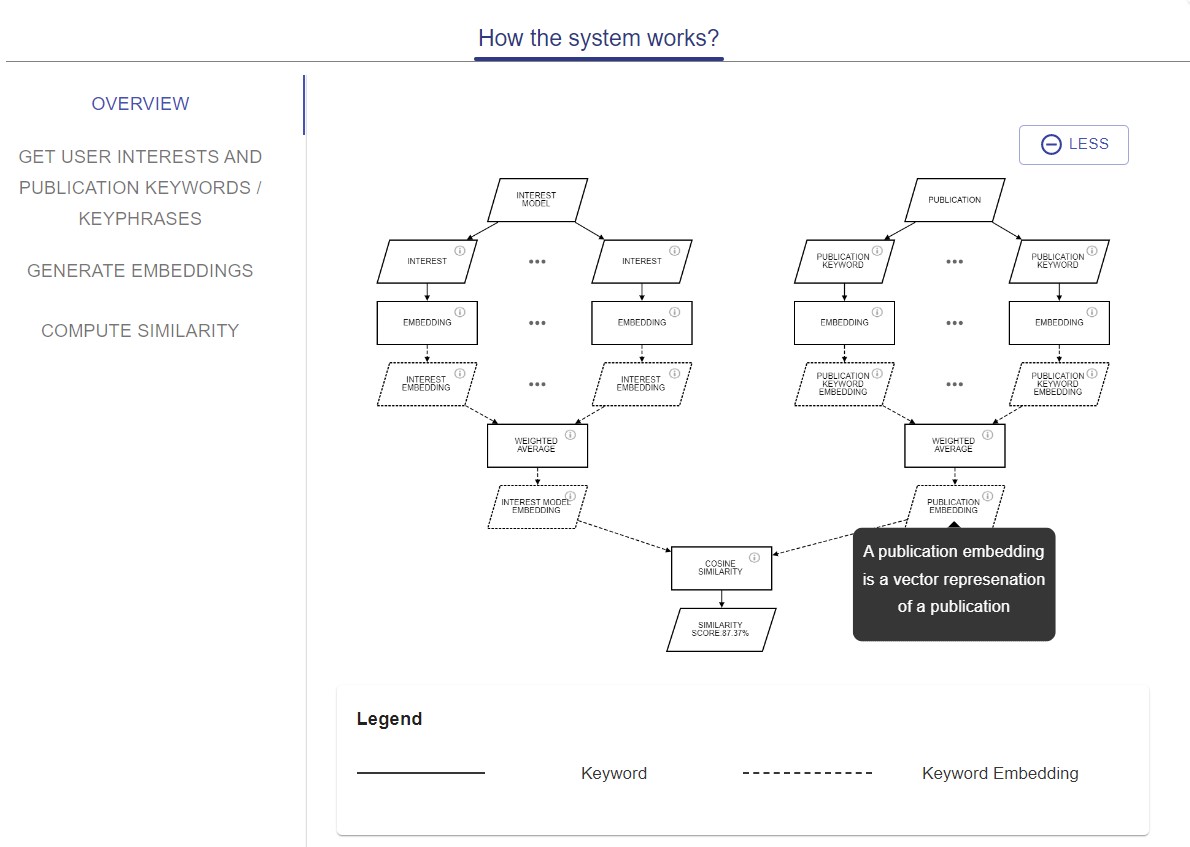}
         \label{fig:how_detail}
     }
        \caption{How intelligibility type.}
        \label{fig:IT3}
\end{figure}

\begin{figure}[htbp]
     \centering

     \subfloat[How – Get user interests and publication keyphrases.]{
         \includegraphics[width=0.8\textwidth]{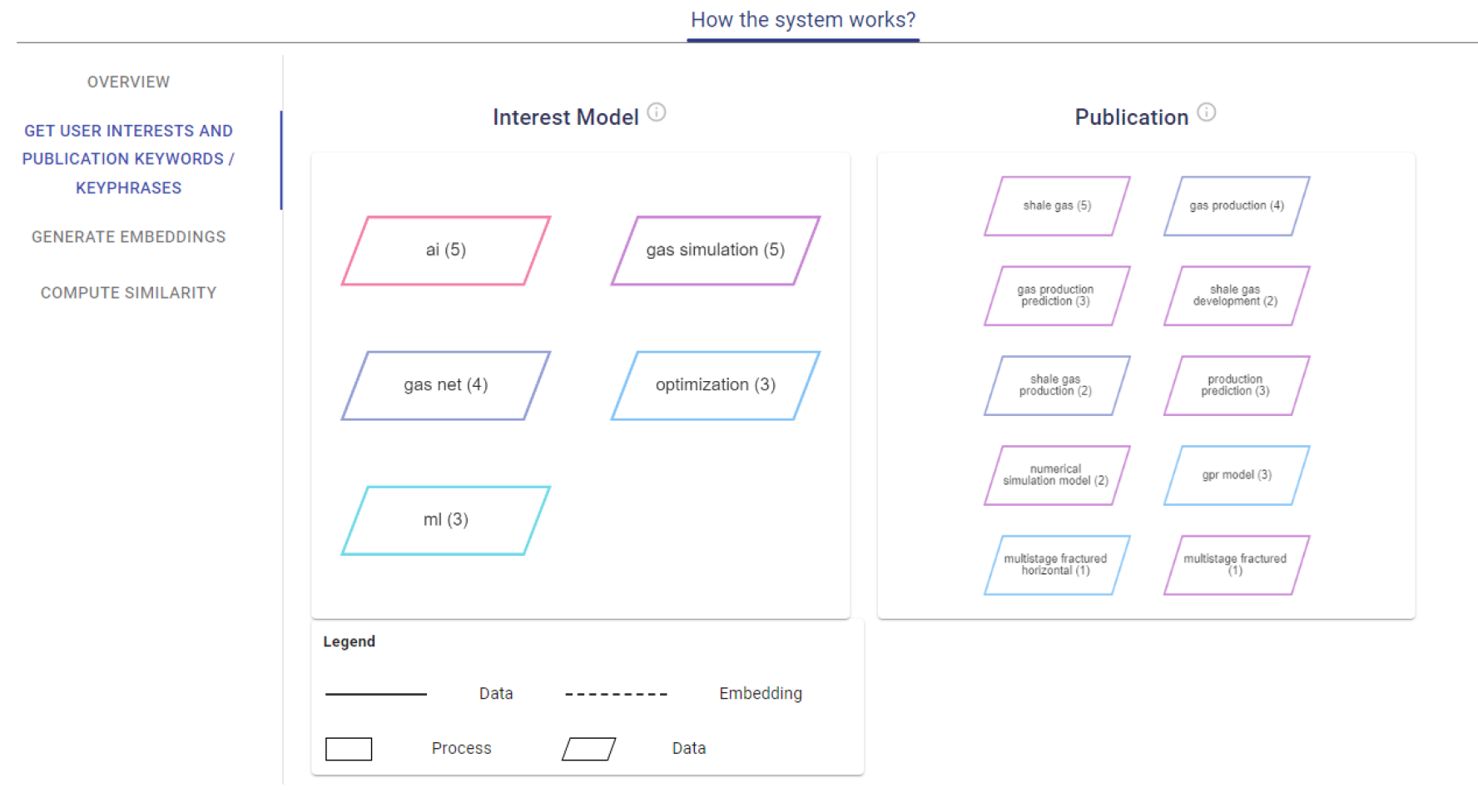}
         \label{fig:interests_and_keywords}
     }
     
     \hfill
     
     \subfloat[How – Generate embeddings.]{
         \includegraphics[width=0.8\textwidth]{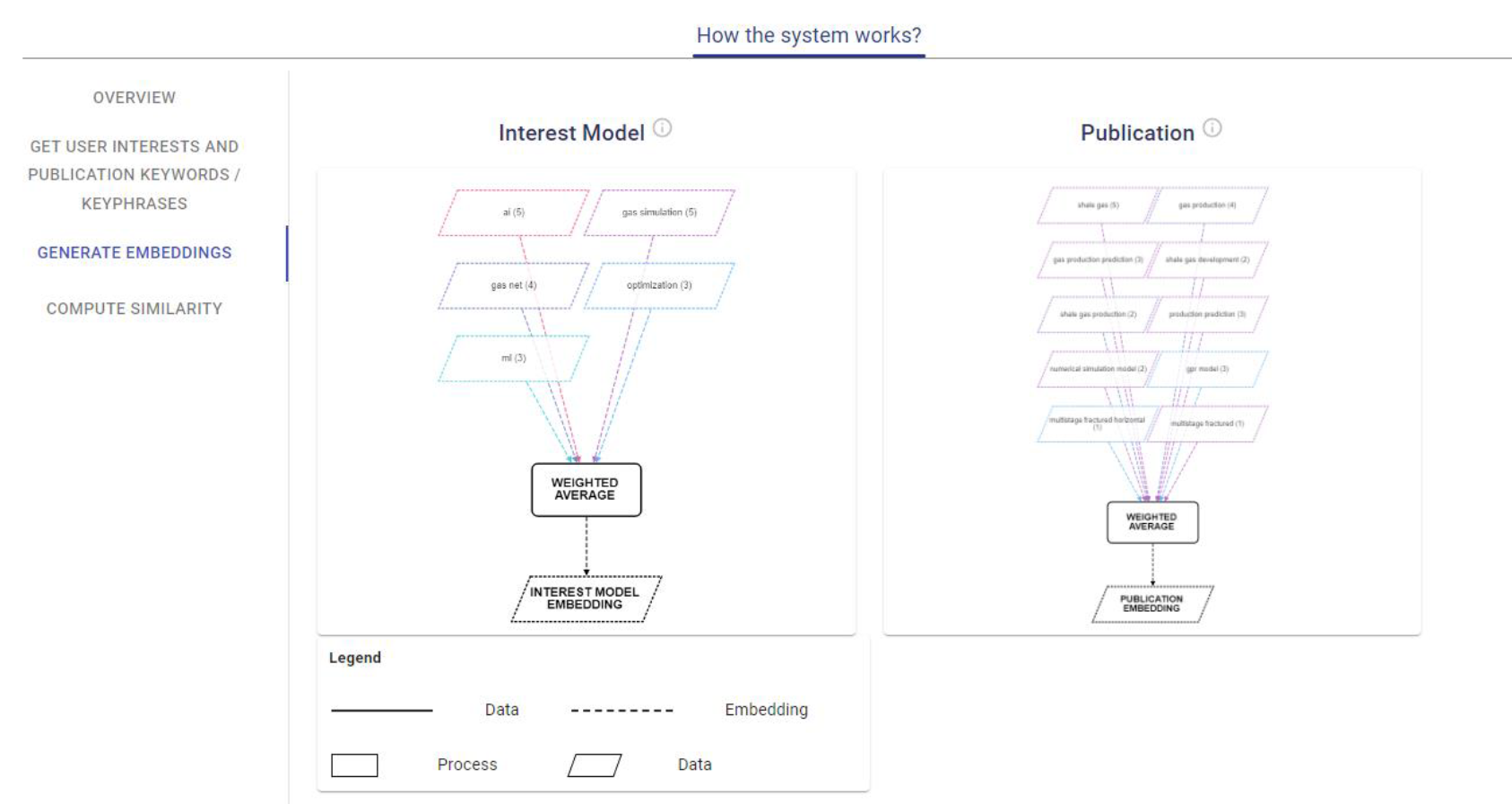}
         \label{fig:Embedding_generation}
     }
     
     \hfill
     
     \subfloat[How – Compute similarity.]{
         \includegraphics[width=0.5\textwidth]{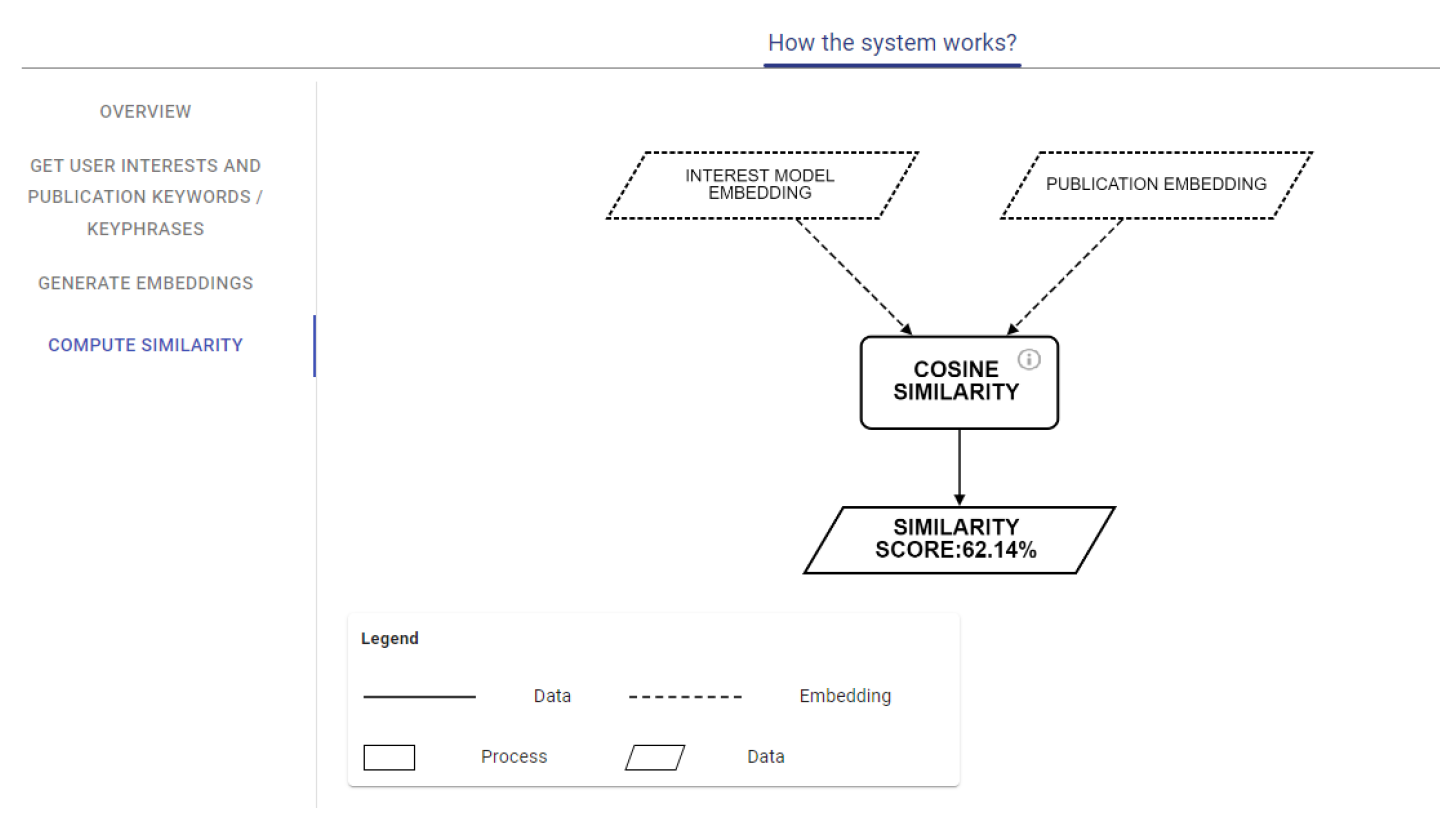}
         \label{fig:sim_computation}
     }
        \caption{Steps of the How intelligibility type.}
        \label{fig:IT4}
\end{figure}

In our earlier works, we followed a human-centered design (HCD) approach \citep{norman2013design} to systematically design \textit{What} and  \textit{What if} explanations \citep{guesmi2022if} as well as \textit{Why} and \textit{How} explanations \citep{guesmi2023info} in the same recommendation domain of scientific publications. In this work, we use these intelligibility types as building blocks of the three explanation levels (see Section \ref{LoD}). The explanatory visualizations related to the different intelligibility types are shown in Figures \ref{fig:IT1},\ref{fig:IT2},\ref{fig:IT3} and \ref{fig:IT4}. 
The \textit{What} explanation reveals to the users what does the system know about them and what kind of information has been used as input for the RS.  For this explanation, we display the user's top five interests with different colors (Figure \ref{fig:what_int}). 
The \textit{What if} explanation presents an exploratory manipulation (i.e., add, remove, change the weights) of the user's interests.  It enables users to test various scenarios and check whether a publication will still be recommended or not after making the adjustments. We use colors to distinguish between the publications’ recommendation statuses to immediately see the impact of the changes users performed on their interest model (Figure \ref{fig:whatif_int}). 

The \textit{Why} explanation justifies why a specific item was recommended. We distinguish between two types of \textit{Why} explanations. In the \textit{Why (abstract)} explanation, when users get a list of publications related to their interests, a similarity score is displayed in the top right corner of each recommended publication box. Moreover, a color band on the left side of the box indicates the similarity score between the current publication and each user's interest. Furthermore, the main keywords of the recommended publication are highlighted in bold format and presented in the color of the most similar interest (Figure \ref{fig:why_abstract}). 
The second type is the \textit{Why (detailed)} explanation which is intended for users who seek further information on the reasons behind providing such recommendations. Through a tag cloud, bar chart, and colors, this explanation shows a more detailed justification of why a certain publication was recommended. Users can hover over each keyword of the recommended publication in the tag cloud to see its similarity to all interests in the bar chart (Figure \ref{fig:why_detail}). 

The \textit{How} explanation targets users willing to know how the underlying RS algorithm works exactly. An overview and the three main steps of the RS algorithm, namely (1) get user interests and publication keyphrases, (2) generate embeddings, and (3) compute similarity, are provided in a navigation panel on the left side of the explanation flowchart (Figure \ref{fig:how_overview}). By hovering over each node in the flowchart, users can get a brief textual description about it. By clicking on the "MORE" button located in the top right corner of the explanation panel, users can expand the flowchart to see more details (Figure \ref{fig:how_detail}). 

Furthermore, by clicking on each step in the left navigation panel, users can stepwise expand the flowchart to see a personalized explanation based on their individual data (Figure \ref{fig:interests_and_keywords}, \ref{fig:Embedding_generation} \& \ref{fig:sim_computation}). 

\subsection{Explanation level of details} \label{LoD} 
The amount of information that should be presented to users in an explanation is still an unsettled question. 
In this work, we are interested in systematically designing explanation with varying level of details with the aim of providing adequate amount of information at each level through manipulating the level of explanation soundness (i.e., nothing but the truth) and completeness (i.e., the whole truth). Similar to \citep{kulesza2013too}, we use the intelligibility types proposed by \citep{lim2009assessing} to define three levels of completeness namely "Low", "Medium", and "High", with more complete explanations including more of these intelligibility types, as suggested by \citep{kulesza2015principles}. The more intelligibility types employed, the more part of the algorithm will be exposed. However, low-completeness (LC) was discarded based on findings from \citep{kulesza2013too} as it was shown to be ineffective and might cause an oversimplification problem. Thus, we decided to proceed with medium-completeness (MC) which explains a significant part of the system, and high-completeness (HC) which explains the whole system. We addressed medium-completeness (MC) with the intelligibility types \textit{What}, \textit{What if}, and \textit{Why} to cover the input and output parts of the RS. As for the high-completeness (HC), besides what we have in MC, we add the \textit{How} intelligibility type to cover all the system's parts (i.e., input, process, output). 

Similar to completeness, we can have three levels of soundness, namely low-soundness (LS), medium-soundness (MS) and high-soundness (HS). The higher the explanation soundness is, the more accurate the explanation is in describing the system algorithm. Low-soundness (LS) explanation could be presented through an abstract justification of why a certain publication was recommended (in our case the \textit{Why (abstract)} explanation). Medium-soundness (MS) explanation should provide more information compared to the LS explanation, but still should not fully explain how the algorithm works (in our case the \textit{Why (detailed)} explanation). The high-soundness (HS) explanation should thoroughly explain the inner working of the RS (in our case the \textit{How} explanation).  

Our goal is to provide explanation with varying level of details based on different combinations of completeness and soundness levels. For each level of completeness, we have three levels of soundness. LC will be neglected as explained above. For MC, three combinations with the three soundness levels can be formed: MCLS, MCMS, and MCHS. MCLS and MCMS will be both used as it was shown in \citep{kulesza2013too} that reducing soundness while preserving completeness will improve the perception of the explanations. MCHS, however, will be neglected for the same reason that was provided in \citep{kulesza2013too}, namely the over-complexity effect. \cite{kulesza2013too} pointed out that increasing completeness alongside soundness mitigates the over-complexity effect. Consequently, for the HS level, only one combination is still possible which is HCHS. Finally, we obtained three possible combinations (i.e., MCLS, MCMS, HCHS). We considered MCLS as \textit{basic explanation}, MCMS as \textit{intermediate explanation}, and HCHS as \textit{advanced explanation}. Table \ref{table:2} summarizes the three explanation levels, alongside the combinations of completeness and soundness and the intelligibility types used at each explanation level. 

\begin{table}[h]
\centering
\begin{tabularx}{\textwidth}{|X|>{\centering\arraybackslash}p{5.5cm}|>{\centering\arraybackslash}p{5.8cm}|}
 \hline
 Explanation level & Level of completeness/soundness & Intelligibility types \\
 \hline
 Basic  & MCLS & \textit{What}, \textit{What if}, \textit{Why (abstract)}  \\ 
 \hline
 Intermediate  & MCMS & \textit{What}, \textit{What if}, \textit{Why (abstract)}, \textit{Why (detailed)}  \\
 \hline
 Advanced  & HCHS & \textit{What}, \textit{What if}, \textit{Why (abstract)}, \textit{Why (detailed)}, \textit{How} \\
 \hline
\end{tabularx}
\caption{Explanation levels: basic, intermediate, and advanced.}
\label{table:2}
\end{table}

 \subsection{User-centered design approach}
We followed a user-centered design approach to systematically design on-demand interactive explanations with varying level of details by involving end-users and consulting them for the evaluation of the explanation prototypes to ensure that their needs and requirements are taken into consideration throughout the design process. The aim of applying a user-centered design methodology was to gradually improve the design of the basic, intermediate, and advanced explanation interfaces, discussed in Section \ref{LoD} and the appropriate arrangement of the intelligibility types, presented in Section \ref{Int_Types}. Next, we outline the user-centered design process of the three explanation levels and their evaluation.

\subsubsection{Prototyping and testing}
We used the different intelligibility types (see Figures \ref{fig:IT1},\ref{fig:IT2},\ref{fig:IT3} and \ref{fig:IT4}) as building blocks to create prototypes for the explanation interface at three different levels of detail (i.e., basic, intermediate, and advanced). For evaluating the explanation interface prototypes, five potential users were involved to test and give feedback on the provided prototypes, as recommended by \citep{nielsen2000five} in the case of qualitative user studies. Our target group are researchers and students who are interested in scientific literature. We recruited five participants (three females) from the local university who were master’s graduates or higher and familiar with information visualization. We asked participants to think aloud when interacting with the prototypes to gain in-depth feedback regarding the layout of the explanation interfaces, the information provided at each explanation level, the placement of the different intelligibility types, the button labels, and the transition between the different explanation levels.

For the basic explanation, the three intelligibility types \textit{What}, \textit{What if}, and \textit{Why (abstract)} are provided in the main explanation interface.  
For the \textit{What} explanation, we generated multiple layouts to determine where users would expect to find the explanation, whether it should be visible or hidden, and if accessible via a button, what would be its label. The purpose of this explanation was to inform users about the data used by the system to provide recommendations, specifically, the top five user's interests extracted from their publications. 
When participants were asked to test this explanation, they suggested that it should be prominently displayed rather than hidden. Further, when we presented users with various prototypes that displayed the list of interests in different places, they preferred to have the explanation at the top of the main page \ref{fig:what_int}. 
The \textit{What if} explanation enables users to manipulate (i.e., add, remove, change the weights) their interests, thus enabling them to explore various scenarios and monitor changes in recommended publications, which are highlighted in different colors on a bar chart (Figure \ref{fig:whatif_int}).
We asked users about their thoughts on the information provided in this explanation, its placement on the explanation interface, and how they access it. They found it useful to have their list of interests displayed within this explanation. Moreover, as this explanation would not be used frequently, they agreed on accessing it on-demand and preferred to have a button with the label "WHAT-IF?" next to the list of interests in the \textit{What} explanation. 
The \textit{Why (abstract)} explanation describes why a specific item was recommended through colors matching between the interests and the keywords in the abstract text of the recommended publication, a similarity score, and a color band showing how much the recommendation is relevant to each interest (Figure \ref{fig:why_abstract}). 
When we asked users about their opinion about how to access this explanation and its placement, they preferred to have this explanation prominent on the explanation interface and shown by default. They also agreed to see this explanation underneath the \textit{What} explanation to easily perceive the color matching between the main keywords
of the recommended publication and the most similar interest.
Finally, we asked users about their thoughts on the amount of information provided in the basic explanation consisting of the \textit{What}, \textit{What if}, and \textit{Why (abstract)}. Participants agreed that seeing the information (i.e., their generated interests) that the RS had at its disposal, and how it used that information to provide recommendations is sufficient to help them understand why certain items are recommended by the system and control the recommendation process, if needed. 

\begin{figure} [h]
  \centering
  \subfloat[Why (detailed) explanation as part of an expandable/collapsible accordion.]{\includegraphics[width=0.4\linewidth]{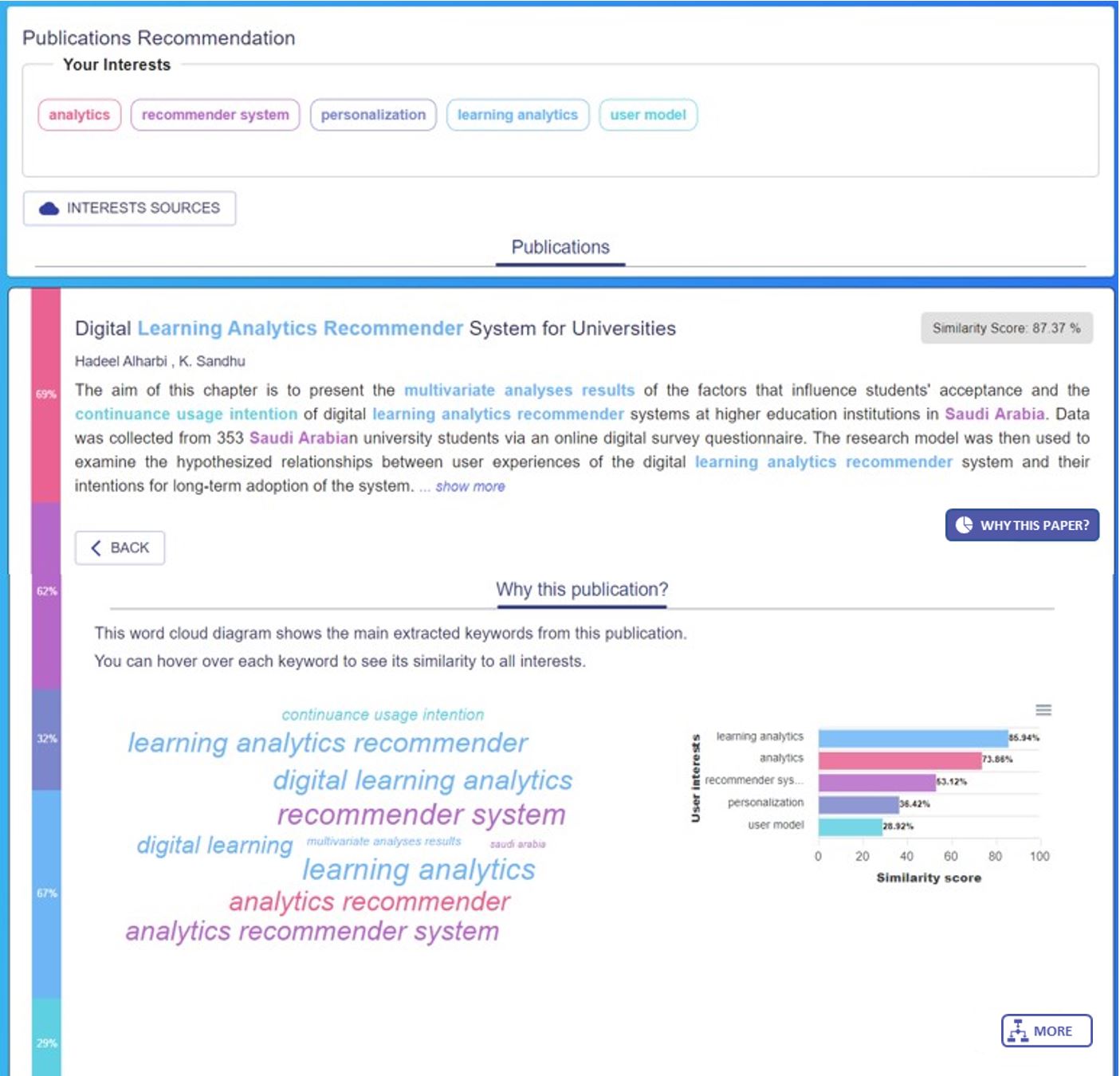}\label{fig:accordion}}
  \hfill
  \subfloat[Why (detailed) explanation as a pop-up modal on top of the publication box.]{\includegraphics[width=0.5\linewidth]{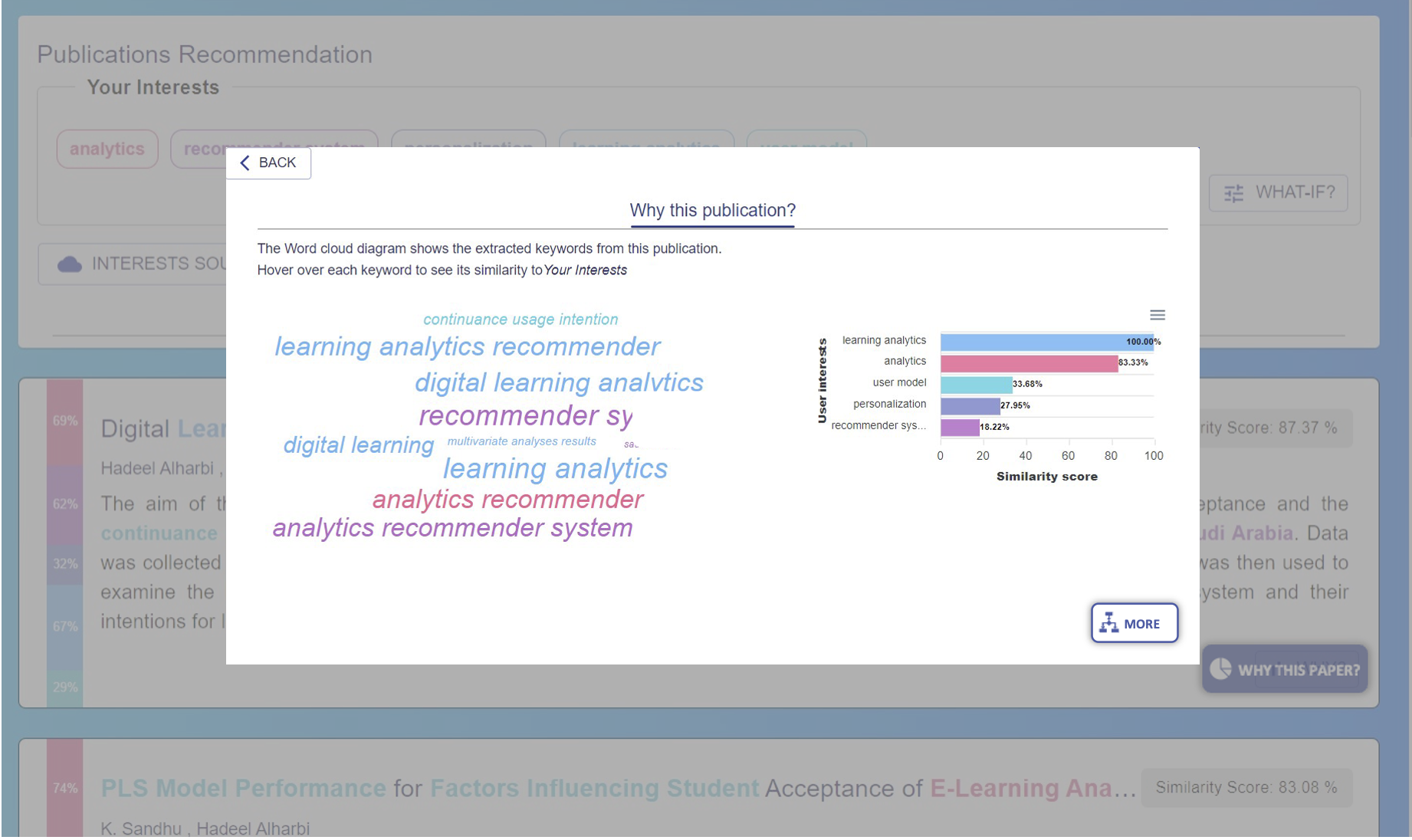}\label{fig:popup}}
  \caption{Prototyping - intermediate explanation.}
  \label{fig:D_P1_Intermediate_explanation}
\end{figure}

The intermediate explanation extends the basic explanation with \textit{Why (detailed)} intelligibility type which describes why a specific item was recommended with more technical details. A word cloud will be displayed on the left side that represents keywords extracted from the publication, and a bar chart on the right side, which appears upon clicking on one of the keywords to show the similarity between the selected keyword and the user's interests. We provided two prototypes to show the \textit{Why (detailed)} explanation on-demand by clicking on a "WHY THIS PAPER?" button within each recommended publication box (Figure \ref{fig:D_P1_Intermediate_explanation}). Two options were provided for the placement of the \textit{Why (detailed)} explanation. In the first option, the publication box will appear as an expandable/collapsible accordion to show the \textit{Why (detailed)} explanation on-demand (Figure \ref{fig:accordion}). The second option will display the explanation as a pop-up modal on top of the publication box (Figure \ref{fig:popup}). 
We asked users about their opinion regarding the transition from the basic to the intermediate explanation, the button labeling, and the placement of the \textit{Why (detailed)} explanation. Users agreed to have the \textit{Why (detailed)} explanation on-demand, thus allowing them to choose whether to see more details or not. They proposed changing the button label from "WHY THIS PAPER?" to "WHY?" to reduce the amount of text and have it consistent with the "WHAT-IF?" button.  
The majority of participants preferred the first option to display the Why (detailed) explanation as part of an expandable/collapsible accordion. This would allow them to view everything on a single page. Most participants opinioned that displaying this explanation as a pop-up modal on top of the publication box would hide important information from the \textit{What}, \textit{What if}, and \textit{Why (abstract)} explanations.

\begin{figure}[h]
\centering
\includegraphics[width=0.6\textwidth]{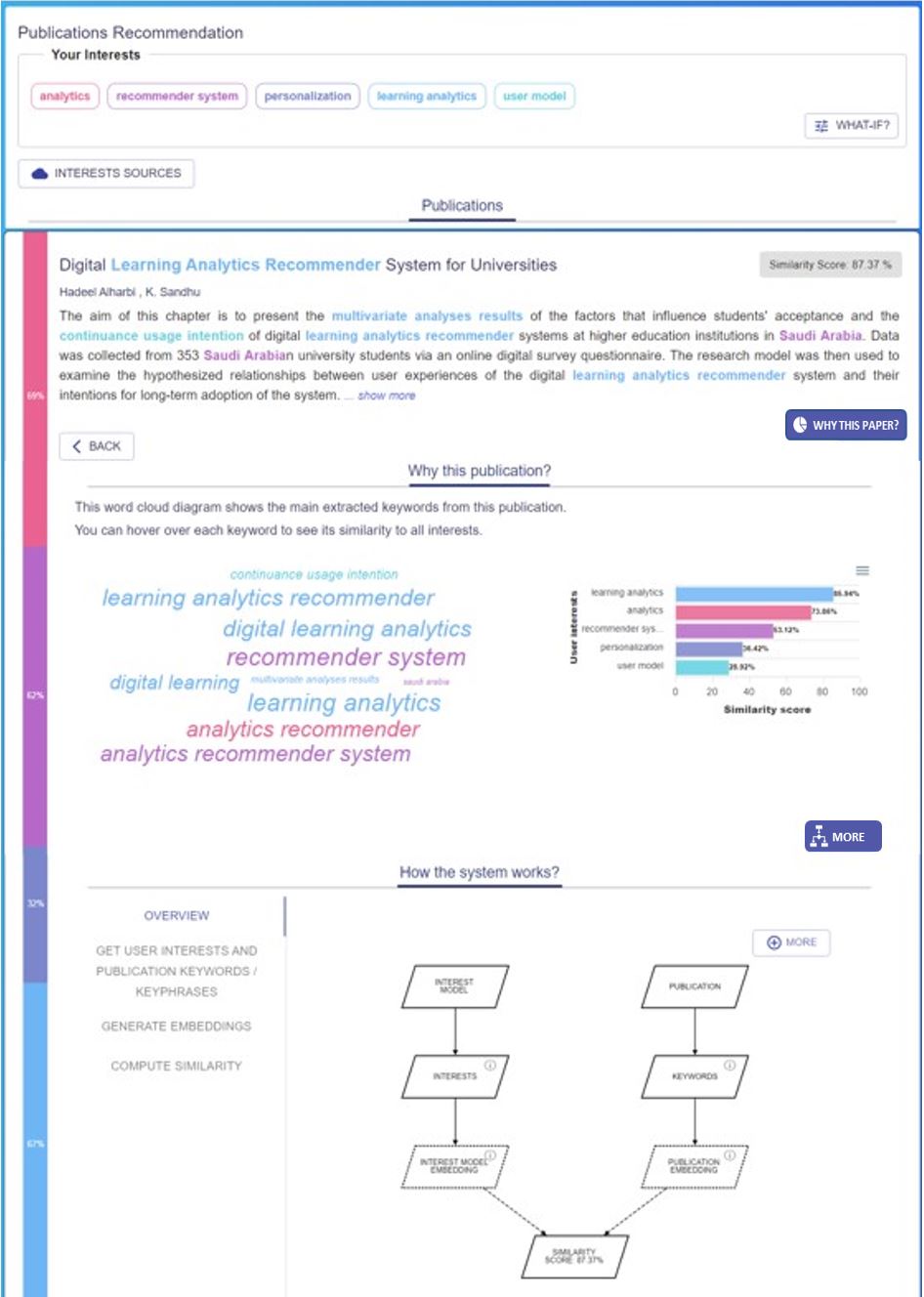}
\caption{Prototyping - advanced explanation.}
\label{fig:D_P1_advanced_explanation}
\end{figure}

At the advanced explanation level, we strive to increase the level of explanation detail by incorporating the \textit{How} intelligibility type in addition to the existing explanations. The \textit{How} explanation employs a flowchart to elucidate the inner working of the system, providing a comprehensive understanding of its operation at different stages (see Figures \ref{fig:how_overview}, \ref{fig:how_detail}, \ref{fig:interests_and_keywords}, \ref{fig:Embedding_generation} \& \ref{fig:sim_computation}). This explanation will be displayed upon clicking on a "MORE" button at the bottom right corner of the \textit{Why (detailed)} explanation box, which will further expand the explanation interface to show the flowchart and the navigation panel (Figure \ref{fig:D_P1_advanced_explanation}). 
We elicited feedback from users regarding their perception of the advanced explanation. We specifically inquired about the transition from the intermediate to the advanced explanation, the positioning of the \textit{How} explanation, and the labeling of the navigation buttons. Participants expressed that they were able to distinguish the different levels of explanation and comprehend the relationship between the \textit{What}, \textit{Why}, and \textit{How} explanations more easily due to the color-coding utilized. Users also found the transition between the explanation levels to be seamless and understandable. Moreover, they liked that the advanced explanation is not shown by default and that they are able to see this explanation when they would need more information about the recommendation process. As for button labeling, they suggested renaming the "MORE" button leading to the advanced explanation to "HOW?" in order to avoid the potential ambiguity of the term "MORE" which could mislead users expecting to see additional information about the \textit{Why (detailed)} explanation. Labeling the button with "HOW?" would reflect more the purpose of the action behind this button, i.e., to show how a recommendation was generated. Furthermore, using the term "HOW?" would keep consistency in the button labeling throughout the whole explanation interface.

 \subsubsection{Final design and implementation}
\begin{figure} [h]
  \centering
  \subfloat[What and Why (abstract) explanation]{\includegraphics[width=0.45\linewidth]{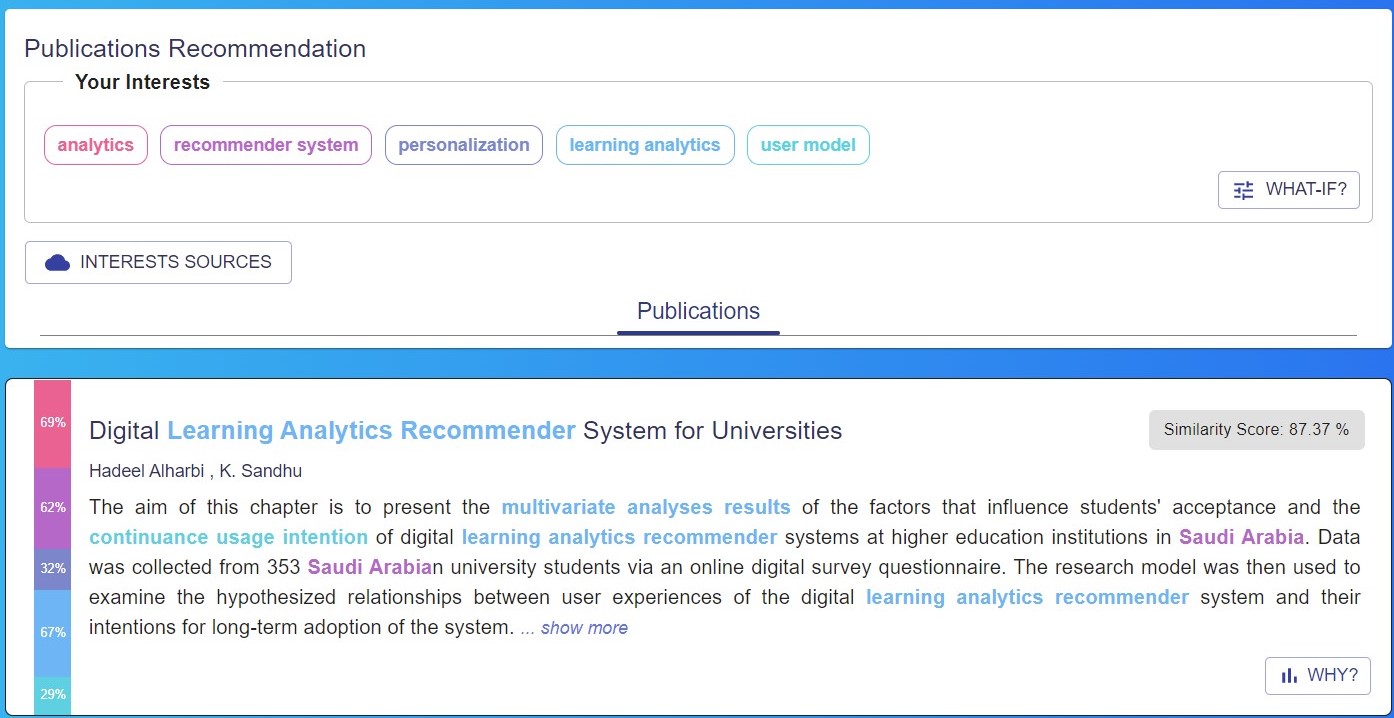}\label{fig:basic_what}}
  \hfill
  \subfloat[What if explanation]{\includegraphics[width=0.45\linewidth]{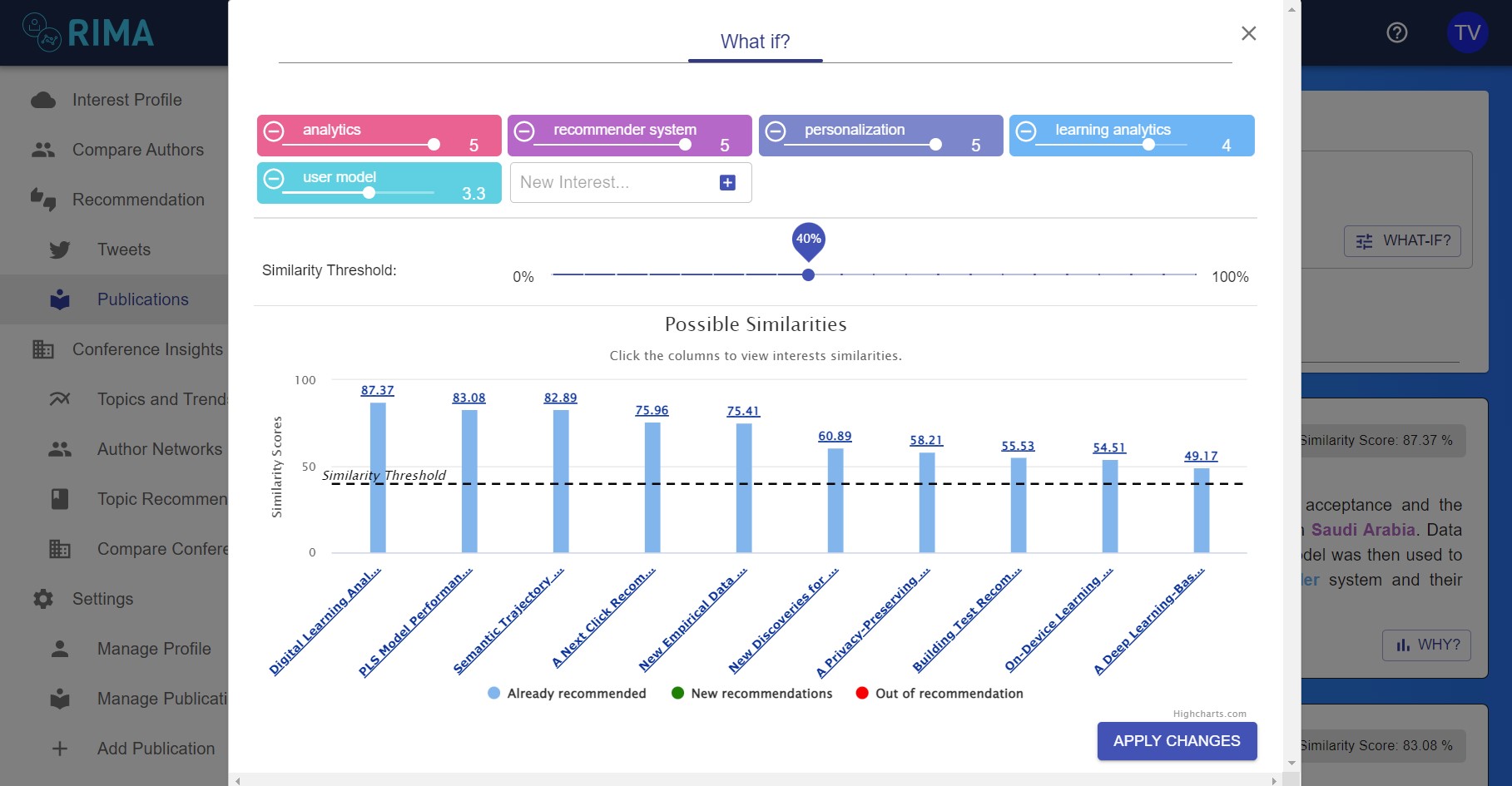}\label{fig:basic_what_if}}
  \caption{Final design - basic explanation}
  \label{fig:basic_explanation}
\end{figure}

\begin{figure} [h]
  \centering
  \subfloat[Intermediate explanation: What, What if, Why (abstract), Why (detailed).]{\includegraphics[width=0.4\linewidth]{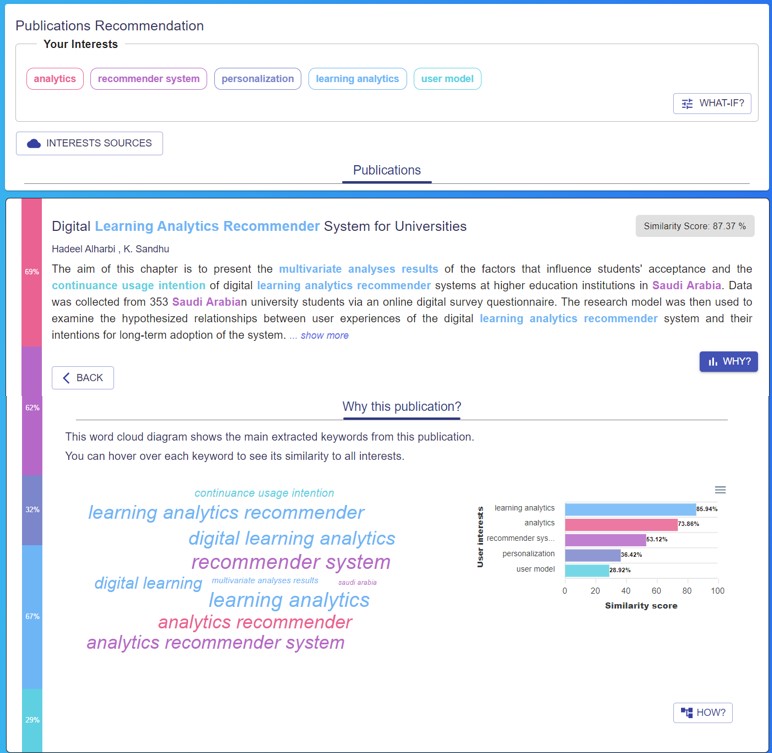}\label{fig:Intermediate_explanation}}
  \hfill
  \subfloat[Advanced explanation: What, What if, Why (abstract), Why (detailed), How.]{\includegraphics[width=0.4\linewidth]{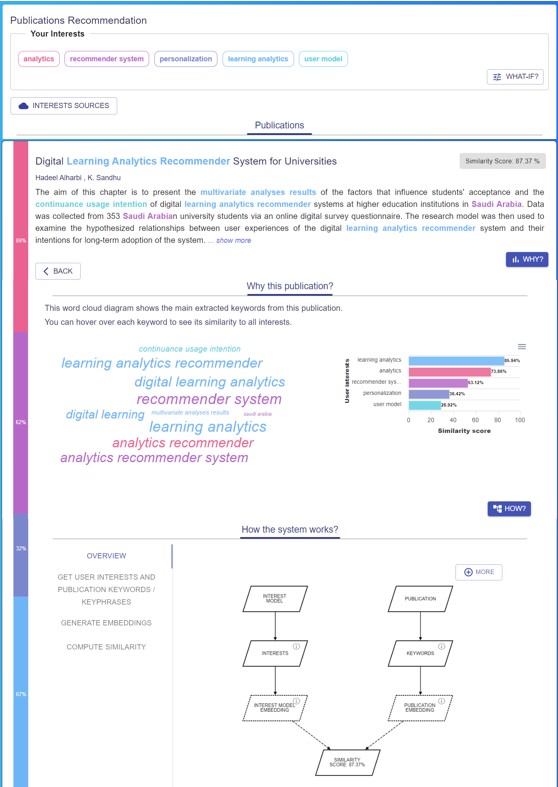}\label{fig:Advanced_explanation}}
  \caption{Final design - intermediate and advanced explanations.}
  \label{fig:}
\end{figure}

Based on the previous iteration and considering the feedback from users, the final explanation prototype was implemented in the RIMA application. Overall, the main suggested improvements included providing a "WHAT-IF?" button next to the list of users' interests at the top of the explanation interface, having the publication box as an expandable/collapsible accordion to show the \textit{Why (detailed)} explanation on-demand, rather than displaying the explanation as a pop-up modal on top of the publication box, and Changing the "WHY THIS PAPER?" and "MORE" buttons to "WHY?" and "HOW?", respectively.  In terms of placement of the different intelligibility types and transition between the different explanation levels, the final explanation interface shows the \textit{What} and \textit{Why (abstract)} intelligibility types by default, while the \textit{What if}, \textit{Why (detailed)}, and \textit{How} intelligibility types are available on-demand. Users can click the "WHAT-IF?" button to access the \textit{What if} explanation, the  
"WHY?" button to move from the basic to the intermediate explanation showing the \textit{Why (detailed)} intelligibility type, and finally the "HOW?" button to move to the advanced explanation with additionally the \textit{How} intelligibility type. 
We used React and several visualization libraries such as Cytoscape and Highcharts to implement the final explanation prototypes. Figure \ref{fig:basic_explanation}, Figure \ref{fig:Intermediate_explanation}, and Figure \ref{fig:Advanced_explanation} show the implemented versions of the basic, intermediate, and advanced explanation, respectively. 

%%%%%%%%%%%%%%%%%%%%%%%%%%%%%%%%%%%%%%%%%%%%%%%%%%
\section{Evaluation}
After systematically designing the explanations with varying level of details and implementing them in the RIMA application, we conducted an online qualitative user study to explore the expectations and attitudes towards our scientific literature RS, considering these various explanations. Since users’ perspectives on and expectations of the explanation level of detail in an explainable RS are inherently subjective and individual, a quantitative approach would be insufficient
for this goal as this approach aims at analyzing empirical data for
predetermined hypotheses. Thus, to explore individual expectations from an explainable RS with varying level of details, we deem a qualitative approach as more appropriate. 

\subsection{Study design}
Researchers and students interested in scientific literature were invited to participate in our study. We first informed participants about the interview conditions, such as the purpose and motivation of the interview, the anonymity of the responses, and the assertion that the data would not be shared with third parties. We also requested permission to record the interview audio in case it was required for the data analysis step that followed. 14 participants (six females) agreed to take part in this study.  Participants were between 20 and 39 years old, where half of them were master’s graduates or higher, and the other half were master’s students. To obtain a diverse sample, participants were recruited with different backgrounds, including nationalities, educational levels, and fields of study. Email or word-of-mouth was used to reach each participant. The ethics motion to conduct the user study was approved by the Ethics Committee of the Department of Computer Science and Applied Cognitive Science of the Faculty of Engineering at the University of Duisburg-Essen. All participants gave their informed consent to participate in the study and to have their interviews recorded. All personally identifiable information was anonymized.

The user study is divided into three sections. Participants were initially given a short introductory video about the RIMA application in general, and another short demo video about the provided interactive explanations in the application. Next, they answered a questionnaire in SoSci Survey \footnote{https://www.soscisurvey.de} which included questions about their demographics and familiarity with RS and visualization. Afterwards, we conducted moderated think-aloud sessions where participants were asked to (1) create an account using their Semantic Scholar ID (users who do not have Semantic Scholar IDs can generate their interest models manually) in order to create their interest models, (2) interact with the application to find relevant publications that fit their interests, and (3) take a closer look at the three explanation levels and try to understand how the RS works. Following a think-aloud approach, the participants were also asked to say anything that comes to their mind during each interaction.
After that, we conducted semi-structured interviews to (1) ask users about their expectations regarding the amount of information provided in each explanation level and if it was sufficient and adequate for them, and (2) gather in-depth feedback related to their attitudes towards the interactive explainable RS. 

The interviews lasted 10 to 15 minutes with the following questions: \textit{
\textbf{(1)} What do you like the most about providing different explanations with varying level of information/details?
\textbf{(2)} What do you like the least about providing different explanations with varying level of information/details?
\textbf{(3)} Which explanation provided you with an adequate amount of information/details? Why?
\textbf{(4)} Which explanation level of information/details (Task 1, 2, 3 / basic, intermediate, advanced) do you prefer (is suitable for you)? Why?
\textbf{(5)} Which explanation level of information/details (Task 1, 2, 3 / basic, intermediate, advanced) is sufficient for you to make a decision (on whether the recommended publications are relevant or not)?
\textbf{(6)} Why / When (in which situation) / How often would you like to use each of the provided explanations?
\textbf{(7)} Do you have any suggestions to improve the controllability of the level of information/details of the explanations provided in the recommender system
\textbf{(8)} How much has the controllability of the level of information influenced your satisfaction with the recommender system?
\textbf{(9)} Which explanation level of information/details (Task 1, 2, 3 / basic, intermediate, advanced) gives you a better sense of transparency of the recommender system? Why?
\textbf{(10)} Which explanation level of information/details (Task 1, 2, 3 / basic, intermediate, advanced) gives you a better sense of trust in the recommender system? Why?
(\textbf{11)} Do you have any suggestions to improve the system?
}

After the semi-structured interviews, participants were also invited to fill out a questionnaire containing questions regarding usability aspects and attitudes towards the RS, based on the ResQue evaluation framework \citep{pu2011user}. To note that by using the ResQue framework, we are not aiming at conducting a quantitative evaluation and generalizing our conclusions, but rather to use participants' answers to the ResQue questionnaire as a starting point to collect their opinions towards our scientific literature RS, considering the three levels of explanation detail, which are then explored in-depth through our qualitative study.

\subsection{Results}

\begin{figure}[h]
\centering
\includegraphics[width=0.9\textwidth]{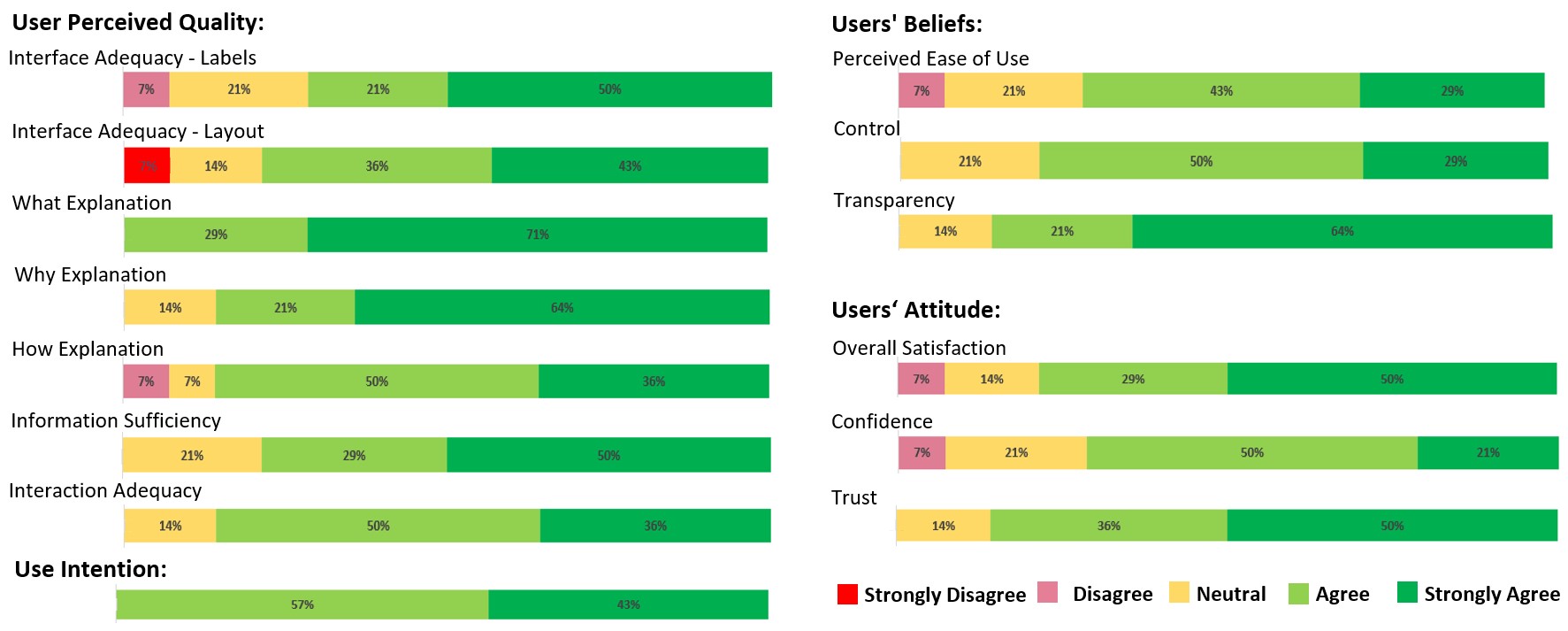}
\caption{Results from the ResQue questionnaire.}
\label{fig:ResQue}
\end{figure}

The results of the ResQue questionnaires are summarized in Figure \ref{fig:ResQue}.
We conducted a qualitative analysis of the moderated think-aloud sessions and the semi-structured interviews to gain further insights into the reasons behind the individual differences in the perception of the RS in terms of explanation with varying level of details. We followed the instruction proposed by \citep{braun2006using} to code the data and identify patterns to organize the codes into meaningful groups. Notes and transcripts of the interview recordings were made for the analysis. 
The analysis was rather deductive as we aimed to find additional explanations to address our research questions. In contrast to inductive (i.e., bottom-up) thematic analysis, which is the data-driven process of coding the data without trying to fit it into a pre-existing coding frame, deductive (i.e., top-down) thematic analysis is an analyst-driven process of coding for a quite specific research question. This form of thematic analysis tends
to provide less a rich description of the data overall, and more a detailed analysis of some aspects of the data \citep{braun2006using}. 
Following a deductive thematic analysis approach, we present the results of the evaluation organized by five themes derived from our two research questions. The first theme is \textit{information level} which is related to our first research question about the amount of provided information at each explanation level. The other four themes for our second research question were adapted from the ResQue framework, namely \textit{user control and personalization}, \textit{transparency and trust}, \textit{satisfaction}, and \textit{overall user experience}.
%%%%%%%%%%%%%%%%%%%%%%%%%%%%%%%%%%%%%%%%%%%
\subsubsection{Information level}
\begin{figure}[h]
\centering
\includegraphics[width=0.65\textwidth]{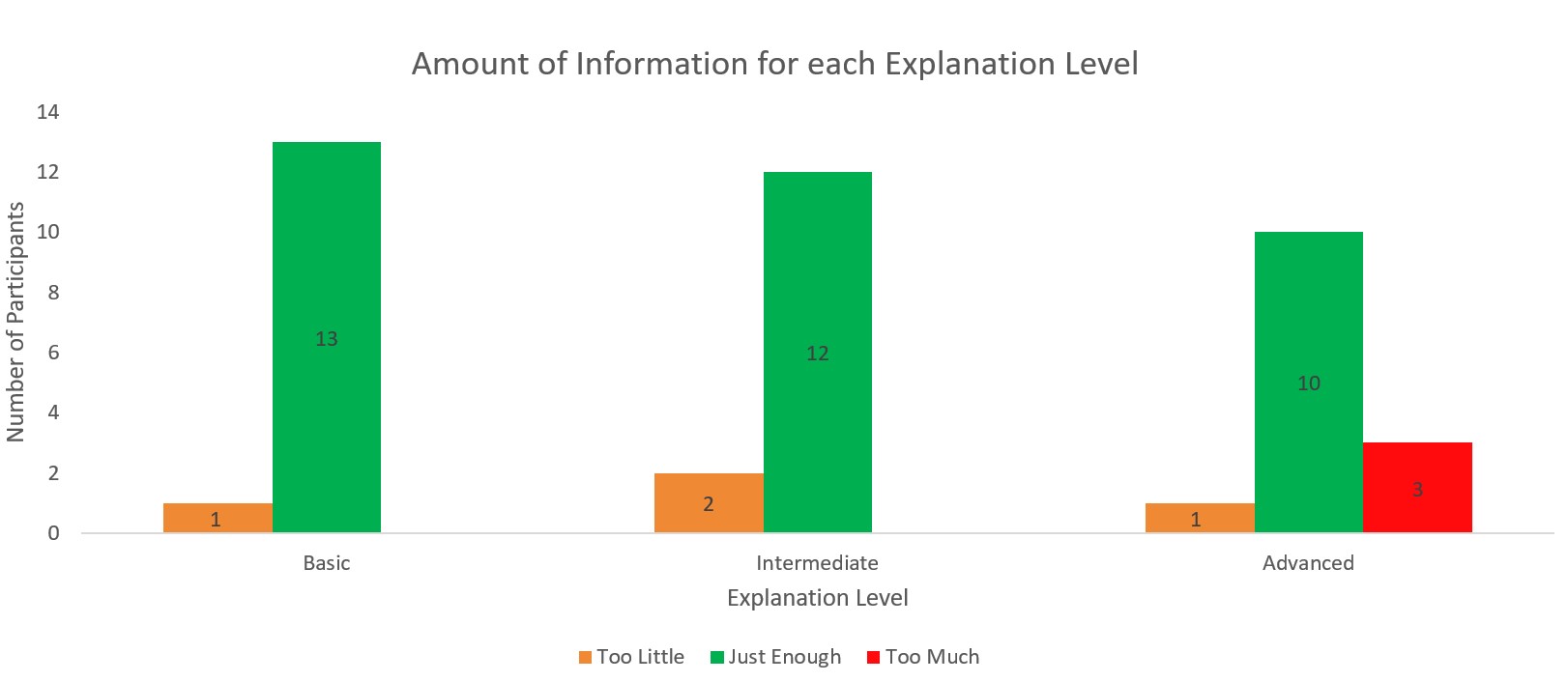}
\caption{Adequacy of the amount of information at the three explanation levels.}
\label{fig:amount_of_info}
\end{figure}

When we asked the participants about the amount of information provided at each explanation level, they had diverse opinions and thoughts (Figure \ref{fig:amount_of_info}). The majority of participants thought the amount of information at the basic level was adequate. However, one participant \textbf{(P14)} stated that the information given was insufficient as he needed more definitions of similarity scores and thresholds. He also mentioned, \textit{"What confused me is the percentages; if you sum them up, they are going over 100\%, so it is not easy to understand what these percentages are"}. Therefore, he assumed more definition is required in this case as well. Likewise, the majority of participants thought that the amount of information provided at the Intermediate level was adequate. Nevertheless, two participants acknowledged that they want to have more information at this level. One of them \textbf{(P3)} wanted more data about the publication content and the placement of the extracted keywords in the publication's text. Another participant \textbf{(P8)} who got an irrelevant publication with a specific keyword mentioned that \textit{"when I hover over the SriLanka word, I still do not know why this keyword has a connection with my interests, even though it is just 2.76\% similar. I would be interested to know how it connects with my interests in this stage"}. For the advanced level, we obtained diverse feedback. Most users believed the amount of the delivered information was sufficient. However, \textbf{P3} assumed that the provided information was too simple: \textit{"I find it basic. The algorithm is basic, if your target user is advanced in this step. It is not what the advanced user does not know already"}. On the contrary, three users believed there was too much information because either there were unknown interests or repeated information. One user assumed that it would be time-consuming to go through each step in detail. For instance, users \textbf{P9} and \textbf{P14} stated, \textit{"I prefer to have the formulas, what the embeddings are. I prefer to see the data as an overview in a tooltip. Try to use simple wordings and show the formulas"}, and user \textbf{P4} reported, \textit{"I felt this is telling me one piece of information in different ways that can be combined all together."}

\subsubsection{User control and personalization}
The majority of the participants expressed positive opinions about the interaction and user control features in the RS (see Figure \ref{fig:ResQue}). Overall, participants had two different perspectives regarding the relationship between controllability and personalization. Some of them saw their control of the explanations as a means to achieve more personalized recommendations, for example, \textbf{P1} said \textit{"I have a feeling of control over recommendations and interests. I can personalize it by adding or removing my interests. So I feel like I can control this and immediately see the result"}. On the other hand, some others considered the control of the level of information given by the explanations as a way of personalizing the explanation itself instead of the RS. For instance, \textbf{P2} stated that \textit{"I like when I can hide something I do not need to see in the explanation. Or I can show it easily. I like this personalization option}". 

Most users expressed positive feedback toward controlling the provided information in the explanations, and they mentioned this feature as the most liked feature in the RS. For instance, \textbf{P13} reported \textit{"It was flexible and I could see anything that I prefer to see. I did not see such a platform so far. It is quite nice"}. Despite some participants assuming that certain explanations are not necessary for them or they would not use them frequently, all participants expressed that the ability to control the information provided by the explanations is a valuable feature and can be helpful. Therefore, they all wanted to keep this option. For instance, \textbf{P14} reported \textit{"It is good to have the possibility to show or hide something, otherwise there is too much information on one page, and you have to scroll down a lot to get to another thing"}.

%BASIC
Particularly, for the basic level of explanations, most users were satisfied with the interaction with and control of this explanation. The \textit{What} explanation was prominent on the explanation interface. For instance, \textbf{P13} mentioned that \textit{"It was easy and interesting to see the data on which the system bases its recommendation process"}. Likewise, the \textit{What If} and \textit{Why} explanations were considered as very helpful. For instance, \textbf{P5} reported that \textit{"I find it interesting. The problem is with other platforms that you cannot easily weigh keywords that are much more important than others. However, sometimes you want to focus on some keywords more than others. This way of personalization is nice"}, and \textbf{P6} stated \textit{"The coloring is quite intuitive"}.
%InTERMEDIATE
Most participants gave positive feedback to the information at the intermediate explanation level. However, they expected to improve the visualizations and controllability at this level, for instance, \textbf{P4} mentioned: \textit{"The most important thing you want to show users is hidden in a tooltip"}, and \textbf{P1} stated: \textit{"The aim of the three lines in the why explanation is not clear, because the bar chart will appear just by hovering over the keywords and we are not able to click on it"}.
%ADVANCED
The advanced level received more diverse feedback. Some users believed that the design of the steps was beneficial because they could follow and learn in-depth how the system operates. For example,  \textbf{P8} mentioned \textit{"It is great that you have here a practical example of how the model works"}, and \textbf{P11} stated \textit{"I think it is cool that you can also see it here visualized, because just showing the model or the process is boring for me. So as an average user, not from IT, I am more interested in seeing some visuals and graphics, but not a complicated model"}. However, we got some feedback regarding the improvement of the way of personalizing the \textit{How} explanation. Many users suggested having the actual presentation of the values or formulas in each step. For instance, \textbf{P8} mentioned that \textit{"Why not keep having the practical numbers here so that persons like me, who are unfamiliar with computer science, could still follow"}, and \textbf{P1} stated: \textit{"Here is the weighted average in each step; you can show the real information maybe instead of a weighted average or actual number for the embeddings, it may help the user"}. On the other hand, some participants found it unnecessary to know how the system works in detail. One of them \textbf{(P2)} mentioned \textit{"I think the more detailed visualizations are unnecessary because everything is completely clear here in the overview"}.
\subsubsection{Transparency and trust} 
\begin{figure}[h]
\centering
\includegraphics[width=0.4\textwidth]{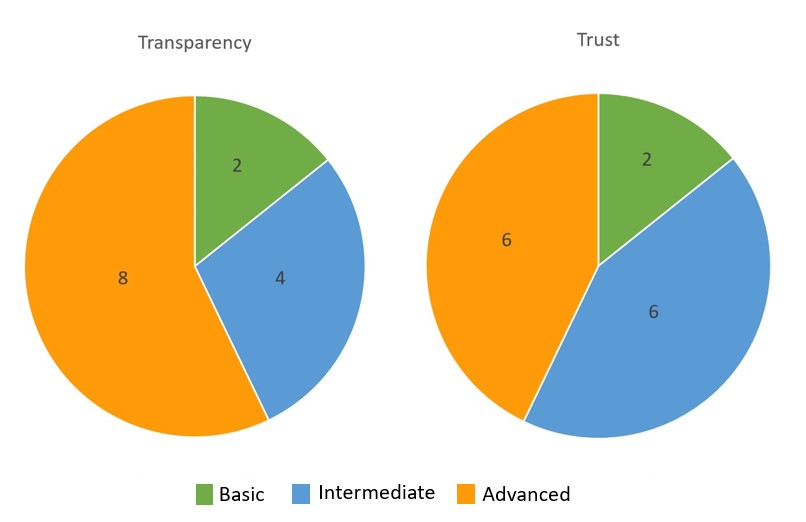}
\caption{Users' perception of transparency and trust at the three levels of explanation detail.}
\label{fig:ResQue_Trans}
\end{figure}

This theme concerns the perception of the three explanation levels in terms of transparency and trust (Figure \ref{fig:ResQue_Trans}). 

In this regard, almost all participants stated that the provided visual explanations had an overall good effect on transparency and trust in the system. Eight participants reported that the system was transparent because of the advanced level of explanation and stated that the system's inner working was evident. For instance, participant \textbf{P1} reported, \textit{"I understood the How explanation, especially going through all steps helped me follow the structure from top to bottom. I can see, follow, and understand where these keywords and colors come from"} and  \textbf{P13} reported, \textit{"There are steps of calculating the similarity score, it is quite obvious and it is nice to have many levels to see step by step what is going on here"}. Four participants expressed that the intermediate level gives them a better sense of transparency, they were either users who are a bit familiar with RS or who want to use the explanations to better control the recommendations. For instance, \textbf{P12} mentioned, \textit{"Intermediate is more transparent for me, maybe for two reasons; the first one is that this is the part that I will rely on to choose the suitable recommendation. The second reason is that I already have the basic background knowledge of RS. So I might not be interested more in how it works, because I already know it"}. Similarly, \textbf{P3} cited, \textit{"I need this transparency as long as I can improve my choices of interests and the publication results, and I think the flow is enough"}. Only two participants assumed that the basic level was enough to understand the RS, and one of them \textbf{P5} stated that \textit{"The basic one was very clear. Even I do not know anything about programming and algorithms. I can very easily understand how the application is working"}.

Regarding the perceived trust in the system, most participants found that the system is trustworthy through the intermediate and advanced levels of explanation. For instance, among the six users who selected the advanced level, one of them \textbf{P10} believed that \textit{"it gives me more assurance of what is happening"}. Likewise, \textbf{P11} stated that \textit{"I can see where it comes from, so I can build my trust on it"}. Six other participants experienced trust at the intermediate level. They assumed that the provided similarity scores would increase trust in the system because \textit{"\textbf{P14}: I see how much these scores, that I had in this cloud are somehow matching my interests"}, and \textit{"\textbf{P11}: I like the idea of seeing the keywords and also showing the similarity score. So it logically makes sense, so it is not randomly chosen, and you can see why it was chosen"}. Two participants stated that the basic level is enough to trust the system. They trust the system as long as the recommended publications are relevant to them. They said that as soon as the platform gives them irrelevant results, they would lose their trust in the system.

\subsubsection{Satisfaction}
Most participants expressed high satisfaction with the RS. According to \citep{tintarev2007survey}, satisfaction can also be measured indirectly, measuring user loyalty. Thus, users’ use intentions can be seen as an indirect measure of loyalty and satisfaction with the system. In this regard, all participants expressed their intention to return to the system or recommend it to their friends (see Figure \ref{fig:ResQue}). This can imply their overall satisfaction with the RS. When we concretely asked about the degree to which the controllability of the level of explanation detail influenced their satisfaction with the RS, all participants agreed that providing various informational levels and the capability to control the explanation level of detail indeed helped them use the system in a more effective way, which increased their satisfaction with the RS. In general, participants expressed their satisfaction in various ways. Some were satisfied with the controllability of the RS, some with the functionality of the explanations, and others with the interaction with the visual explanations.

At the basic level, most of the users were satisfied with the controllability of the RS. Specifically, they were satisfied with the \textit{What if} interactions and agreed that it was beneficial and unique. In this regard, \textbf{P2} cited \textit{"I am totally satisfied with that What if explanation because it led me to find the desired paper"}. Likewise, the coloring part in the \textit{What} and \textit{Why} explanation was satisfying for all participants as they believed it was intuitive to see the connections between their interests and the recommended publications. However, these explanations were not convincing for some users as in some recommended publications, some keywords were the same as users' interests but they were not highlighted.
For the intermediate level, the functionality provided in this level was quite satisfying for the participants. For example, \textbf{P3} mentioned that \textit{"There was a time that I found a word somewhere. And I do not know, what is it. Then I saw it is related to me. This helped me to know more about such words and their relations to my interests"}. Nonetheless, we got some suggestions to improve the layout of this explanation, such as \textit{"When I saw the why button, I did not expect to see such a graph but to see the source of it. You have to choose another visualization here"} \textbf{(P4)}, and \textit{"At first this blank space at the right side was strange"} \textbf{(P12)}. 
Regarding the advanced level, we received mixed feedback. Most participants found the general idea, explanation functionality, controllability, and visualization useful and satisfying. Both lay and experienced users found the option of having a personalized visual representation of the different steps of the recommendation algorithm as a flowchart to be helpful. Lay users felt that having a personalized visualization of the RS inner working was appealing. Otherwise, they would find it uninteresting and may not fully understand what is being explained. For instance, \textbf{P4} mentioned that \textit{"I liked this visualization much more than others. It is attractive and helpful"} and \textbf{P14} stated that \textit{"It is responsive and useful"}. The personalized visualizations of the RS algorithm were also used by users who were computer scientists to troubleshoot the system and identify more effective search parameters. However, some participants found the advanced explanation overwhelming. Others wanted to have more information about the technical terms used in this explanation (e.g., embeddings). 

\subsubsection{Overall user experience}
\begin{figure}[h]
\centering
\includegraphics[width=0.25\textwidth]{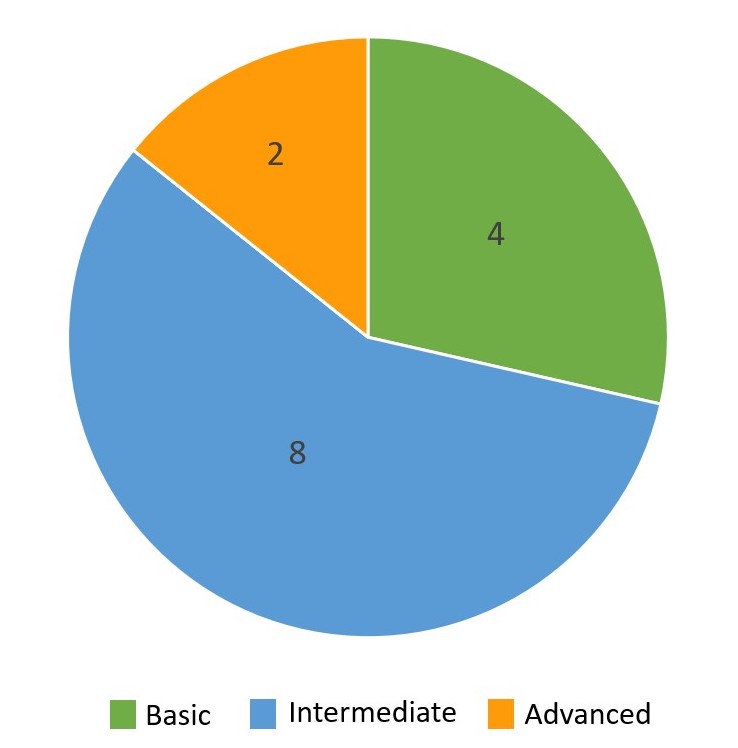}
\caption{Users' preferred explanation level of detail for making a decision.} \label{fig:user_preference}
\end{figure}

The system’s overall mean rating is relatively high, indicating that the overall user experience is positive toward having explanations with different levels of detail. Figure \ref{fig:ResQue} further shows that aspects related to users' beliefs and attitudes, such as user control, transparency, and trust scored higher than the overall mean rating. On the other hand, the system’s interface design and ease of use received relatively lower ratings. For this theme, we also gathered feedback concerning the effectiveness of the explanations and the situations where each explanation could be used. The intermediate explanation was perceived as the most effective level of detail by eight participants (Figure \ref{fig:user_preference}). They assumed that this level was enough for them to make a decision on whether the recommended publication was relevant to them or not. Four participants (respectively two participants) mentioned the advanced explanation (respectively the basic explanation) as the most helpful explanation for making a decision. Regarding the usage context and frequency for each explanation, for the basic level, the majority of the participants stated that they will use the \textit{What if} explanation most frequently in order to change their interests and consequently receive better recommendations. The \textit{What} and \textit{Why} explanations represented by the coloring part and similarity scores were the second most frequent explanations the users tend to pay attention to. Regarding the intermediate level, we discovered five primary situations where the users would use the \textit{Why} explanation. First, some users consulted this explanation when they got irrelevant publication recommendations, as mentioned by \textbf{P1}, \textbf{P4}, \textbf{P9}, and \textbf{P13}. Second, some users used this explanation to discover new keywords from the relevant publications when they receive accurate results, as stated by \textbf{P6} and \textbf{P14}. Third, others preferred to use this explanation to adjust their initial interests in RS to get better results, as noted by \textbf{P3} and \textbf{P5}. Fourth, when the publication abstract is too short \textbf{(P7)}. Fifth, when a user did not know the meaning of an extracted keyword \textbf{(P3)}. Concerning the advanced level, most users said they would use it at least once but not frequently. The majority of users stated that it is an interesting option and can be helpful. For instance, \textbf{P6} said \textit{"At a higher level, I want to know how the system works. I will click on the HOW button, but I would say not so frequently"}. Two participants (\textbf{P2} and \textbf{P7}) said that they would look at the \textit{How} explanation only if they had difficulty understanding the other explanations.

%------------------------------------------------------

\section{Discussion}
The subsequent sections are organized regarding our two main research questions: Which and how much information should be provided at each explanation level of detail? \textbf{(RQ1)} Can providing interactive explanation with varying level of details positively affect the perception of the explainable RS in terms of (1) user control \& personalization, (2) transparency \& trust, (3) user satisfaction, and (4) user experience? \textbf{(RQ2)}

\subsection{Amount of information at each explanation level of detail}
We addressed our first research question by systematically designing three different explanations with varying level of details in an explainable
scientific literature RS.  We started by collecting requirements from the literature to divide the information into different explanation levels, based on the principles of completeness and soundness \citep{kulesza2013too}. We then systematically designed and implemented three explanation levels of detail (basic, intermediate, advanced) with different combinations of completeness and soundness levels, based on \textit{What}, \textit{What if}, \textit{Why}, and \textit{How} intelligibility types.   
According to our evaluation results, most users indicated that the provided explanations with varying level of information were adequate, helpful, and user-friendly. In terms of the provided amount of information at each level, most users agreed that there was enough information for the intended level of detail. However, some lay users sought more details about each technical phrase at a higher level. On the other hand, we observed that users who were computer scientists or familiar with RS tended not to go into the detail of the advanced level because they felt they already knew how the algorithm works. 
This finding suggests a relationship between the users’ background knowledge and the needed amount of information for each level. This is in line with the findings in e.g., \citep{chatti2022more,kouki2019personalized,millecamp2019,szymanski2021visual}, showing that personal characteristics have an effect on the perception of and interaction with explanations. Additionally, as mentioned by \citep{kulesza2013too}, completeness should be enhanced alongside soundness to avoid overwhelming the users. In other words, if the underlying algorithm is fully disclosed, there should be more explanations. Overall, our results show that manipulating the level of explanation soundness and completeness, as well as using different intelligibility types (e.g., \textit{What}, \textit{What if}, \textit{Why}, \textit{How}) to vary the level of completeness is an effective mechanism to systematically design explanations with varying level of details, with the right amount of information to be revealed at each explanation level.  

\subsection{Effects of interactive explanations with varying level of details}
In our second research question, we investigated the degree to which the interactive explainable RS with varying level of details can affect the perceived control, transparency, trust, and satisfaction of its users. 

\subsubsection{User control and personalization}
Our aim was to provide an interactive explanation interface to help users control and personalize the explanation process, based on their individual needs and preferences. By interacting with the explanation interface, the users are now able to control and personalize the explanation by changing the explanation to answer \textit{What}, \textit{What if}, \textit{Why}, and/or \textit{How} questions, showing or hiding the explanation, seeing different levels of explanation detail, and changing the explanation viewpoint to focus on the input, process, and/or output.  
Most users experienced control over the RS and its provided explanations. All participants tried to control the RS by modifying their interests and receiving personalized recommended publications, as well as controlling the explanations by selecting the appropriate level of detail they wish to see. Both lay and experienced participants agreed that having the explanations on-demand was a great idea since it offered them a sense of control over the quantity of information that the explanation should provide. 
Consequently, we argue that, regardless of the users’ background knowledge, allowing users to control and personalize the provided explanations should be an integral feature in any explainable RS.  

\subsubsection{Transparency and trust}
Our study provides evidence that user interaction with and control of the explanation process can improve the RS transparency. This confirms findings in previous studies on explainable recommendation which pointed out that control and transparency are interdependent (e.g., \citep{eiband2018bringing,tintarev2015explaining}). This is also in line with studies from different other application domains, including XAI \citep{cheng2019explaining, sokol2020one, krause2016interacting, kulesza2015principles}, human-centered AI \citep{shneiderman2020bridging}, interactive machine learning \citep{amershi2014power}, interactive recommendation \citep{he2016interactive,jugovac2017interacting,tsai2017providing}, and visual analytics \citep{spinner2019explainer} which showed that human control can also contribute to increased transparency of AI and decision making systems. 

Different levels of explanation detail would lead to different levels of RS transparency. Here, it is important to differentiate between objective transparency and user-perceived transparency. On the one hand, objective transparency means that the RS reveals the underlying algorithm of the recommendations either by explaining it or justifying it in case of high complexity of the algorithm. On the other hand, user-perceived transparency is based on the users’ subjective opinion about how good the system is capable of explaining its recommendations \citep{gedikli2014}. In general, it can be assumed that a higher level of explanation detail increases the system’s objective transparency but is also associated with a risk of reducing the user-perceived transparency, depending on the users’ background knowledge. We observed a similar pattern in our study. More than half of the users perceived transparency at an advanced level of explanation detail, which means that high completeness and high soundness levels benefit system transparency. As these steps were designed to be on-demand, most users mentioned that it was a useful feature that helped them follow and understand the explanations better. However, some participants had difficulty understanding technical terms at the advanced level, as they were unfamiliar with the technical aspects in RS. These users perceived higher transparency rather at a low or intermediate level of explanation detail. 
In our study, most users regardless of their background knowledge concurred that the RS was perceived as transparent. One reason we deem responsible for the high level of user-perceived transparency among the majority of participants is that they were able to find in the system the appropriate explanation level of detail that best fits their needs and preferences. This suggests that empowering users to take control of the explanation process and change the level of explanation detail according to their needs has the potential to improve user-perceived RS transparency. 

For system trust, we obtained nearly the same findings as transparency, which was in line with the work of \citep{tintarev2011designing}, who found that transparency increases user trust. Our findings revealed that most participants identified intermediate and advanced levels as trustworthy. Some participants mentioned that the intermediate explanation justifies the recommended publication, which is why it is reliable, while others believed that revealing the system’s algorithm gave them the confidence to rely on the system. Moreover, the personalized information provided at all explanation levels gave them a sense of trust in the RS. Additionally, two users chose the basic level because they assume that they trust the system as long as it provides useful recommendations. 
This outcome is also consistent with \citep{tintarev2011designing}, who noted that user trust in the RS might be influenced by how accurate the recommendation algorithm is. As the intermediate and advanced explanations achieved the highest levels of transparency and trust, we found that the level of completeness and soundness should be correlated and parallel to help users better understand and trust the system. This finding is also in line with the work of \citep{kulesza2013too}.

\subsubsection{Satisfaction and user experience}
The wide agreement among participants that their interaction with and control of explanations had a positive impact on their satisfaction with the RS shows a potential positive correlation between user control and satisfaction/user experience. Our results confirm the findings by \citep{pu2011user} who noted that user control weighs heavily on the overall user experience with the RS. Furthermore, several recent studies from the human-AI interaction (HAII) community evidenced the positive effects of user control on satisfaction/user experience. For instance, \cite{sundar2020rise} stresses that the two key features of HAII, namely user awareness and user control are the hallmarks of successful user experience with personalization services.

With regard to user experience, our study results show that user-centered aspects, such as feeling of control, transparency, and trust can hide the negative effects resulting from usability issues. Further, the intermediate explanation was perceived as the most effective level of detail to make a decision. This suggests that, if an explainable RS only provides a single static explanation, the focus should rather be on providing an intermediate explanation with enough completeness and soundness to meet the demands of a larger user group. This is in line with the suggestion provided by \citep{kizilcec2016much} who concluded that designing for effectiveness requires balanced interface transparency, i.e., “not too little and not too much”. Finally, participants in our study gave different reasons why and when they want to see which explanation (e.g., to explore, discover, scrutinize, understand, debug, or just out of curiosity). This implies that the users' goal and context influence their preferences towards the explanation intelligibility type and level of detail. Thus, interactive explanations with different levels of detail would be a flexible and effective solution to help users achieve different explanation goals in different contexts, which can contribute to increased satisfaction and user experience. In sum, our work provides qualitative evidence that providing interactive explanation with varying level of details can also increase user satisfaction and experience with the RS. 

%------------------------------------------------------
\section{Limitations}
Our work is also subjected to several limitations. 
Firstly, we identify some limitations related to explanation design. We designed and implemented a subset of possible intelligibility types (i.e., \textit{What}, \textit{What if}, \textit{Why}, and \textit{How}). Other intelligibility types, such as \textit{Why not} (contrastive explanation) and \textit{How to} (counterfactual explanation) are not commonly used in the recommendation domain and are beyond the scope of this paper. Supporting other intelligibility types and exploring their combinations to provide explanations at different levels of detail may lead to different results. 

Moreover, the selection of the recommendation model (i.e., content-based) in our work may bring biases in our results about the effects of interactive explanations with different levels of detail. Our findings may not generalize to other recommendation models (e.g., deep learning-based, graph-based). Deep learning-based recommendation introduces other critical challenges with respect to explanation design. 
Deep learning models are inherently complex, making it often difficult to provide sound explanations that truly reflect the real mechanism that generated the recommendations \citep{zhang2020explainable}. Further, providing advanced explanations of deep learning recommendation models is only beneficial for expert users to help them diagnose and refine the underlying models. This would restrict the design space of explanations with varying levels of detail in a deep learning-based RS. On the other hand, graph-based models inherently keep interpretability in their recommendations, as they can provide more intuitive explanations by extracting relation paths between the target user and the recommended item over the graph \citep{guo2020survey}. This would help in designing and generating appropriate path-wise explanations at varying level of details, for different end users. It is therefore important to explore in the future work whether the difference in the recommendation model could lead to a different explanation design and investigate the extent to which the choice of the recommendation model can influence the perceived explanation quality.

There are also some limitations in the user study. We conducted a qualitative user study with 14 participants. Therefore, the results of the study should be interpreted with caution and cannot be generalized. A quantitative user study with a larger sample would probably have yielded more significant and reliable results. Furthermore, we performed this analysis in a single domain. It must be verified whether our findings transfer to domains beyond scientific literature RS. It would be for example very interesting to look into the effects of providing interactive explanation with varying level of details in a music RS, 
where factors, such as exploration, variety, and novelty appear to play a crucial role in users' perception of transparency and satisfaction \citep{liang2023promoting,millecamp2019,jin2018effects,kouki2019personalized}.

%------------------------------------------------------
\section{Conclusion and future work}
In this paper, we aimed to shed light on two aspects that remain under-explored in the literature on explainable recommendation, namely explanation with varying level of details and interactive explanation. To this end, we systematically designed interactive on-demand explanation with three levels of information detail (basic, intermediate, and advanced) and implemented them in the transparent Recommendation and Interest Modeling Application (RIMA). Applying a qualitative approach, we found evidence that handing over control to users to select their appropriate explanation level of detail can meet the demands of users with different needs, preferences, and goals and consequently can have positive effects on key aspects in explainable recommendation, including transparency, trust, satisfaction, and user experience. This work provides insights for the development of interactive explanation interfaces in RS and contributes to a richer understanding of why and how to design interactive explanation with varying level of details in a systematic and theoretically-sound manner. While we are aware that our results are based on one particular RS and the results cannot be generalized, we are confident that they pose valuable anchor points for the current and future design of interactive explanation interfaces in RS. In future work, we plan to design other possible intelligibility types (e.g., \textit{Why not} and \textit{How to}) and explore other combinations of intelligibility types to provide explanations at the three levels of detail. Further, we plan to validate our findings through quantitative research to investigate in more depth the effects of following a user-centered, interactive explanation approach on the perception of and interaction with explainable recommendation, with different user groups and in different contexts. Our qualitative findings represent a solid base for hypotheses and confirmatory studies on a large user basis. Further, we will investigate the effects of personal characteristics on the perception of interactive explanation with varying level of details in explainable RS. 

%---------------------%

\section*{Acknowledgement(s)}
This work was partly funded by the German Research Foundation (DFG) under grant No. GRK 2167, Research Training Group “User-Centred Social Media”. 
%We acknowledge the support from the Open Access Publication Fund of the University of Duisburg-Essen.

\section*{Notes on contributor(s)}
Conceptualization, M.G., M.A.C. and T.V.; Methodology, M.G. and M.A.C.; Validation, M.A.C.; Software, T.V. and S.J.; Writing—original draft preparation, M.G. and T.V.; Writing—review and editing, M.G. and M.A.C.; Visualization, M.G., T.V., S.J., Q.U.A, R.A. and C.S.; Supervision, M.A.C. All authors have read and agreed to the published version of the manuscript.

\section*{Institutional Review Board Statement}
The study was conducted according to the guidelines of the Declaration of Helsinki, and approved by the Ethics Committee of the Department of Computer Science and Applied Cognitive Science of the Faculty of Engineering at the University of Duisburg-Essen.

\section*{Informed Consent Statement}
Informed consent was obtained from all subjects involved in the study.

\section*{Data availability statement}
The interview data presented in this study are not publicly available due to ethical and privacy restrictions.

\section*{Disclosure statement}
No potential conflict of interest was reported by the author(s).

\section*{Funding}
The authors thank the German Research Foundation (DFG) for partly funding this work under grant No. GRK 2167, Research Training Group “User-Centred Social Media”.

% \section*{Data availability statement}

% An unnumbered section, e.g.\ \verb"\section*{Nomenclature}" (or \verb"\section*{Notation}"), may be included if required, before any Notes or References.

% \section*{Notes}

% An unnumbered `Notes' section may be included before the References (if using the \verb"endnotes" package, use the command \verb"\theendnotes" where the notes are to appear, instead of creating a \verb"\section*").

\bibliographystyle{apacite}
\bibliography{interactapasample}

\section*{About the authors}
\textbf{Mouadh Guesmi} \textit{is a research assistant and PhD student in the Social Computing Group, Department of Computer Science and Applied Cognitive Science at the University of Duisburg-Essen, Germany. His research interests include human-computer interaction, user modeling and personalization, explainable recommender systems, and information visualization.}

\hfill

\textbf{Mohamed Amine Chatti} \textit{is a professor of computer science and head of the Social Computing Group in the Department of Computer Science and Applied Cognitive Science at the University of Duisburg-Essen, Germany. His research focuses on human-centered learning analytics, personalization, recommender systems, trustworthy human-AI interaction, and intelligent explanation interfaces.}

\hfill

\textbf{Shoeb Joarder} \textit{is a research assistant and PhD student in the Social Computing Group, Department of Computer Science and Applied Cognitive Science at the University of Duisburg-Essen, Germany. His main research areas involve human-centered learning analytics, visual analytics, and big data.}

\hfill

\textbf{Qurat Ul Ain} \textit{is a research assistant and PhD student in the Social Computing Group, Department of Computer Science and Applied Cognitive Science at the University of Duisburg-Essen, Germany. Her main research interests include human-computer interaction and explainable recommender systems.}

\hfill

\textbf{Rawaa Alatrash} \textit{is a research assistant and PhD student in the Social Computing Group, Department of Computer Science and Applied Cognitive Science at the University of Duisburg-Essen, Germany. Her research interests include recommender systems, natural language processing, graph neural networks, and user modeling.} 

\hfill

\textbf{Clara Siepmann} \textit{is a research assistant and PhD student in the Social Computing Group, Department of Computer Science and Applied Cognitive Science at the University of Duisburg-Essen, Germany. Her research interests include human-computer interaction, explainable recommender systems, and the relationship between trust and transparency in intelligent systems.}

\hfill

\textbf{Tannaz Vahidi} \textit{has a Master’s degree in computer engineering from the Faculty of Engineering at the University of Duisburg-Essen, Germany.}

%\appendix
\end{document}